\documentclass[a4paper,11pt]{article}
\pdfoutput=1 
\usepackage{jheppub} 

\usepackage[T1]{fontenc} 
\usepackage[utf8]{inputenc}
\usepackage{graphicx,amssymb,amsmath}
\usepackage[outdir=./images/]{epstopdf}
\usepackage{bbold}
\usepackage{booktabs}
\usepackage{slashed}
\usepackage{setspace}
\usepackage{subcaption}
\usepackage{enumitem}
\usepackage{xcolor}
\usepackage{multirow}

\newcommand{\trnsp}{\mathsf{T}}

\newcommand{\ii}{\mathrm{i}}
\newcommand{\D}{\mathrm{d}}
\newcommand{\erw}[1]{\left \langle #1 \right \rangle}
\newcommand{\sN}{\mathcal{N}}
\newcommand{\cN}{N}
\newcommand{\cC}{\mathcal{C}}
\newcommand{\id}{\mathbb{1}}

\DeclareMathOperator{\tr}{tr}
\DeclareMathOperator{\Pf}{Pf}
\DeclareMathOperator{\Sign}{Sign}

\definecolor{darkteal}{HTML}{008080}
\definecolor{b14}{HTML}{9400D3}
\definecolor{b155}{HTML}{009E73}
\definecolor{b17}{HTML}{56B4E9}
\definecolor{b40}{HTML}{E69F00}
\definecolor{b60}{HTML}{0072B2}
\definecolor{b80}{HTML}{E51E10}
\definecolor{b100}{HTML}{000000}

\newcommand{\inputFig}[1]{\includegraphics{#1.pdf}}

\graphicspath{{./images/},{./imagesPDF/}}

\title{\boldmath Mass spectrum of $2$-dimensional $\mathcal{N}=(2,2)$ super
Yang-Mills theory on the lattice}

\author[a]{D. August,}
\author[a]{M. Steinhauser,}
\author[a]{B.~H. Wellegehausen,}
\author[a]{A. Wipf}

\affiliation[a]{Friedrich-Schiller University Jena, Theoretisch-Physikalisches Institut, Germany}

\emailAdd{daniel.august@uni-jena.de}
\emailAdd{marc.steinhauser@uni-jena.de}
\emailAdd{bjoern.wellegehausen@uni-jena.de}
\emailAdd{wipf@tpi.uni-jena.de}

\abstract{
In the present work we analyse $\mathcal{N}=(2,2)$ supersymmetric Yang-Mills 
(SYM) theory with gauge group $SU(2)$
in two dimensions by means of lattice simulations. The theory arises as 
dimensional reduction of $\mathcal{N}=1$ SYM theory
in four dimensions. 
As in other gauge theories with extended supersymmetry,
the classical scalar potential has flat directions which
may destabilize numerical simulations. In addition,
the fermion determinant need not be positive and this sign-problem
may cause further problems in a stochastic treatment.
We demonstrate that $\mathcal{N}=(2,2)$ super Yang-Mills theory 
has actually no sign problem and that the flat directions are 
lifted and thus stabilized by quantum corrections. Only the 
bare mass of the scalars experience a finite additive renormalization
in this finite theory. On various lattices with different lattice constants
we determine the scalar masses and hopping parameters for which 
the supersymmetry violating terms are minimal. By studying four
Ward identities and by monitoring the $\pi$-mass we show that 
supersymmetry is indeed restored in the continuum limit. In the second part 
we calculate the masses of the low-lying bound states. We find that 
in the infinite-volume and supersymmetric continuum limit the 
Veneziano-Yankielowicz super-multiplet becomes massless and the
Farrar-Gabadadze-Schwetz super-multiplet decouples from the theory.
In addition, we estimate the masses of the excited mesons in the
Veneziano-Yankielowicz multiplet. We observe that the gluino-glueballs
have comparable masses to the excited mesons. 
}

\begin{document}

\maketitle
\flushbottom


\section{Introduction}

Many extensions of the standard model of particle physics make use of 
supersymmetry in order to cure well-known flaws of the standard model, 
as for instance the hierarchy problem. Some of the additional particles 
of supersymmetric (susy) gauge theories may be identified as dark matter 
particles in the universe.  Since no additional particles have been observed 
in experiments up to now it is of utmost interest to investigate the spectrum 
of susy gauge theories, in particular in the strongly coupled
regime. The most simple supersymmetric gauge 
theories are probably the $\sN=1$ Super-Yang-Mills (SYM) theories
with gauge groups $SU(\cN)$. These are supersymmetric extensions of $SU(\cN)$ 
Yang-Mills theories \cite{Salam:1974ig,Ferrara:1974pu}. For $SU(3)$ 
the bosonic sector is identical to that of QCD. It describes
the gluons of strong interaction in interaction with their superpartners, 
the gluinos. The gluinos are Majorana fermions transforming in the adjoint representation
of the gauge group.
Like in QCD, the theory is asymptotically free and it is expected, that the gluons 
and gluinos are confined in colorless bound states. 
But differently from one-flavor QCD, the $U(1)_\text{A}$ chiral symmetry is anomalously
broken only to the discrete subgroup $\mathbb{Z}_{2\cN}$. At low temperatures this 
symmetry is further  broken spontaneously to $\mathbb{Z}_2$ by the formation 
of a gluino condensate and thus gives rise to $\cN$  physically equivalent vacua \cite{Amati:1988ft}.
\newline
The  SYM theory has a richer spectrum of colour-blind bound states than QCD 
since the gluinos are in the adjoint representation. Beside (adjoint) mesons, 
baryons and  glueballs, hybrid bound states 
of gluons and gluinos  are expected to show up in the low energy spectrum.
Implementing symmetries and anomalies of the theory, low energy effective 
actions have been proposed \cite{Veneziano:1982ah,Farrar:1998rm,Farrar:1997fn} 
describing the supersymmetric spectrum of bound states. Thereby the chiral multiplet
containing the adjoint f- and $\eta$-meson is extended to a super-multiplet
by a  gluino-glueball. A second multiplet contains a $0^{+}$ glueball, 
a $0^-$ glueball and in addition a gluino-glueball. The low-energy effective action
depends on free parameters and hence it is not clear which multiplet 
is the lighter one. Various arguments were given for both scenarios, 
see \cite{Veneziano:1982ah,Farrar:1998rm,Farrar:1997fn,Feo:2004mr}. 
Another difficulty stems from the fact, that for every state in the first 
multiplet there exists a state in the second multiplet with the same 
quantum numbers. This mixing of states may lead to
an even more complex multiplet structure.

Similarly as QCD the $\sN=1$ SYM theory is strongly coupled 
at low energies and non-perturbative methods are necessary to investigate 
its mass spectrum. We simulate the theory on a discrete spacetime lattice. 
This is a non-trivial task since a lattice regularisation breaks supersymmetry explicitly. This can be seen  from the susy algebra
\begin{equation*}
\left\{\mathcal{Q},\mathcal{Q}\right\}\propto P_\mu,
\end{equation*}
where $\mathcal{Q}$ is a generator of supersymmetry and the $P_\mu$ generate
translations in space and time. Since a discrete lattice does not admit
arbitrary small translations, we can not preserve the full supersymmetry 
on a lattice, similar to chiral symmetry. In order to recover both symmetries
in the continuum limit, certain parameters have to be fine-tuned, making 
simulations more expensive. Fortunately for $\sN=1$ SYM theory,
the only relevant operator that breaks supersymmetry (softly) is a 
non-vanishing gluino condensate which at the same time breaks
chiral symmetry. Thus it suffices to restore
chiral symmetry in the continuum limit to recover supersymmetry \cite{Curci:1986sm},
making chiral Ginsparg-Wilson fermions the preferred choice
\cite{Giedt:2008xm,Endres:2009yp,Kim:2011fw}. Unfortunately chiral fermions are
computationally very expensive such that it seems to be more efficient to fine-tune the 
bare gluino mass parameter of Wilson fermions. For the gauge group $SU(2)$ with 
Wilson fermions, the theory has been extensively investigated by the 
DESY-M\"unster collaboration 
\cite{Montvay:1997ak,Campos:1999du,Farchioni:2001wx,Montvay:2001aj,Munster:2014cja,Bergner:2014saa,Bergner:2014dua,Bergner:2015adz}. 
Their results confirm the formation of the predicted super-multiplets and reveal, 
that the glueballs are heavier than the mesons. Simulations for the gauge group 
$SU(3)$ are underway \cite{Ali:2018dnd,Steinhauser:2017xqc}.
\newline
Another strategy is to look at the dimensionally reduced model, namely 
$\sN=(2,2)$ SYM theory in two dimensions. By calculating the mass spectrum 
of this related and simpler model we should get further insights
into the four-dimensional model. The two-dimensional super-renormalizable
descendant of the four-dimensional theory allows for larger lattices and
much better statistics. This will lead to a mass spectrum with less
statistical errors than in four dimensions.
\newline
A first numerical simulation of the two-dimensional model was presented in
\cite{Suzuki:2005dx,Fukaya:2007ci}, where the dimensional reduction was done 
for the lattice theory with compact link variables. Accordingly the scalar fields in
the reduced model appear in the exponent of the compact link variables. 
In the simulation the quenched configurations were reweighted with the 
Pfaffian. Because of large (statistical) errors the results for
Ward identities were inconclusive.
\newline
Apart from being a descendant of SYM theory in
four dimensions, the $\sN=(2,2)$ theory in two dimensions 
has further interesting properties. Theoretical arguments
\cite{Witten:1995im,Fukaya:2006mg} and
numerical calculations based on a discretized light cone quantization
\cite{Antonuccio:1998mq,Harada:2004ck}, both suggest
massless states in the physical spectrum. This 
massless super-multiplet is not seen in four dimensions. 
Furthermore, it has been conjectured that dynamical susy breaking may occur in the theory \cite{Hori:2006dk}. 
Recent lattice results for the vacuum energy however show no sign of susy breaking \cite{Catterall:2017xox}.
\newline
Analogous to the two-dimensional $\sN=(2,2)$ Wess-Zumino model \cite{Kastner:2008zc},
the two-dimensional $\sN=(2,2)$ SYM theory admits a conserved and nilpotent 
supercharge. This is possible because there are
four supercharges from which one can build \emph{one} nilpotent
supercharge $\mathcal{Q}$. On a lattice only the subalgebra generated 
by  nilpotent supercharges can be realized.
Several $\mathcal{Q}$-exact lattice models were proposed
\cite{Cohen:2003qw,Catterall:2004np,Sugino:2003yb}. 
All these models suffer from the following problem: Usually one can
expand the link variables as $U_\mu=\mathbb{1}+\ii a A_\mu+\cdots$,
in which case we expect an unique vacuum state. This is not the case
in all three models proposed and thus one expects an ambiguous continuum
limit. In the models in \cite{Cohen:2003qw,Catterall:2004np} the 
problem is solved by adding the susy-breaking 
term $\mu^2\,\text{tr}\left(U^\dagger U-\mathbb{1}\right)^2$ to the Lagrangian, which 
dynamically picks a unique vacuum state. 
In the limit $\mu\to 0$, supersymmetry is recovered
in this construction. In contrast, by deforming the model \cite{Sugino:2003yb} 
the unphysical vacuum states can be removed without breaking
the nilpotent supersymmetry explicitly \cite{Matsuura:2014pua}.
Several numerical investigations show the restoration of the full susy
(not only the nilpotent one)
\cite{Catterall:2006jw,Catterall:2008dv,Kanamori:2008bk,Suzuki:2007jt,Kanamori:2008yy,Kanamori:2007yx,
Kadoh:2009rw}. 
The relations between these models were investigated in \cite{Takimi:2007nn,Damgaard:2007eh,Damgaard:2007xi,Unsal:2006qp}. 
For a more detailed overview see the reviews
\cite{Kaplan:2003uh,Giedt:2006pd,Catterall:2009it,Joseph:2011xy,Bergner:2016sbv}.
\newline
Two-dimensional continuum gauge theories have less dynamical
degrees of freedom than four-dimensional ones and thus we may expect that
topology of the (Euclidean) spacetime becomes more important. 
In our work we use periodic lattices which discretize a two-torus.
In the works \cite{Matsuura:2014kha,Kamata:2016xmu,Kamata:2016rqi} 
different lattices with other spacetimes were scrutinized.
In particular a generalized topological twisting on
generic Riemann surfaces in two dimensions \cite{Matsuura:2014kha}
has been considered. The authors revealed the connection of 
the sign problem, which is absent on the torus, to the $U(1)_\text{A}$ anomaly.
With a so called compensator the sign problem can be solved 
on Riemann surfaces with genus $\neq 1$. Ward identities and the
$U(1)_\text{A}$ anomaly -- the latter is intimately related to the zero modes of the 
Dirac operator -- have been looked at.
\newline


The paper is organized as follows: In section \ref{sec::Model} we introduce 
the $\mathcal{N}=(2,2)$ theory, discuss its continuum properties and in particular the expected 
particle spectrum. There is only one relevant operator $\tr\phi^2$ that needs to 
be fine-tuned  to recover susy in the continuum limit. The corresponding 
mass-parameter is calculated to one-loop order. To investigate the restoration of 
susy we derive three independent Ward identities. In section 
\ref{sec::Lattice} we introduce our lattice formulation with Wilson fermions 
and discuss some technical points like the fermion sign problem, potentially flat 
directions of the effective potential and fine-tuning of the bare parameters.
In \cite{Hanada:2009hq,Hanada:2010qg} it was argued that it is important to control flat directions of the scalar potential. We shall see in our simulations that the flat directions are lifted and we observe no instabilities in the scalar 
subsector. Furthermore the model has no sign problem in the simulations.
Since susy is broken at finite lattice spacing, the Ward identities are 
not fulfilled. The additional contributions at finite lattice spacing are 
discussed in section \ref{sec::Ward} for the gauge group $SU(2)$,
together with our simulation results  
concerning the restoration of supersymmetry in the continuum and thermodynamic limit. 
In section \ref{sec::Spectrum} we present our accurate results 
for the masses of the low lying bound states. 
One super-multiplet becomes massless in the 
thermodynamic and supersymmetric limit and a second super-multiplet
decouples from the theory. In addition we see a massive super-multiplet of
excited states. At the end we present our conclusions in section
\ref{sec::conclusions}.


\section{$\sN=(2,2)$ SYM theory in two dimensions}
\label{sec::Model}

\noindent
In this section we will derive $\sN=(2,2)$ supersymmetric Yang-Mills (SYM) theory
in two dimensions by a dimensional 
torus-reduction from $\sN=1$ SYM theory in four dimensions. 
To recall this reduction is useful since there
is a one-to-one correspondence between the $\sN=1$ super-multiplets 
in four dimensions and the $\sN=(2,2)$ super-multiplets in two dimensions.
We expect that related super-multiplets have the same length since
the length can only change when supersymmetry
is (partially) broken or the members of a super-multiplet become 
massless. Thus we may expect that bound states in the two-dimensional theory
arrange in super-multiplets corresponding to
super-multiplets in the four-dimensional theory. 
Note that the assignment of spins in a super-multiplet may change
during the reduction. This happens for the vector super-multiplet but not 
for the chiral super-multiplet. But the mass spectrum may
change, even if there is a one-to-one assignment of super-multiplets.

We begin with reviewing some relevant properties of the
four-dimensional theory \cite{Salam:1974ig,Ferrara:1974pu}. 
The action is given by
\begin{equation}
	S=\int \D^4x\, \tr \left(-\frac{1}{4}F_{MN}F^{MN}+\frac{\ii}{2}\bar{\lambda}\,
\Gamma^M D_M \lambda\right),	\label{Model 4d}
\end{equation}
where capital indices $M,N$ assume the values $0,1,2,3$, the 
matrices $\Gamma^M$ build an irreducible representation of the 
four-dimensional Clifford algebra and  $F_{MN}$ is the field strength tensor
\begin{equation}
	F_{MN}=\partial_M A_N - \partial_N A_M -\ii\,g \left[A_M,A_N\right]
\end{equation}
with gauge potential $A_M$ in the adjoint representation of the 
gauge group $SU(\cN)$. The gauge potential and Majorana-field are components
of the same super-field such that $\lambda$ transforms under the 
adjoint representation as well. Hence, the covariant derivative of the 
Majorana fermion is
\begin{equation}
D_M\lambda=\partial_M \lambda -\ii\,g\left[A_M,\lambda\right].
\end{equation}
The action \eqref{Model 4d} is invariant under the on-shell
supersymmetry transformations
\begin{equation}
\delta_\varepsilon A_\mu=\ii \bar\varepsilon\,\Gamma_M\lambda,\;\;
\delta_\varepsilon\lambda=\ii F^{MN}\Sigma_{MN}\,\varepsilon,\;\;
\delta_\varepsilon\bar\lambda=-\ii \bar\varepsilon\, F^{MN}\Sigma_{MN}
\label{SUSY Trafo 4d}
\end{equation}
with $[\Gamma_M,\Gamma_N]=4\ii\, \Sigma_{MN}$. These transformations
are generated by $\bar\varepsilon Q$, where $\varepsilon$ is a 
constant anticommuting Majorana-valued parameter
and the $\{Q^\alpha\}$ are the four components of the 
Majorana-valued supercharge $Q$.
The Majorana condition relates the four entries of a spinor according to
$\lambda=\lambda_\mathrm{c}=\mathcal{C}\bar\lambda^\trnsp$, 
where $\mathcal{C}$ is a charge conjugation matrix.

The action is also invariant under global $U(1)_\text{A}$ 
transformations
\begin{equation}
	\lambda \rightarrow e^{\ii\alpha\,\Gamma_5} \lambda\,,\qquad
	\Gamma_5=\ii\,\Gamma^0\Gamma^1\Gamma^2\Gamma^3\,.
\end{equation}
In the quantum theory, this chiral symmetry is broken down to $\mathbb{Z}_{2\cN}$ 
via instantons. If a chiral condensate $\erw{ \bar{\lambda}\lambda}\neq 0$ 
forms, it is further broken spontaneously to $\mathbb{Z}_{2}$
\begin{equation}
 U(1)_A \stackrel{\text{instantons}}{\longrightarrow} \mathbb{Z}_{2\cN} 
 \stackrel{\erw{\bar{\lambda}\lambda}}{\to} \mathbb{Z}_{2}\,.
\end{equation}
The $\cN$ physically equivalent vacua are related by 
the discrete chiral rotations
\begin{equation}
 \lambda \to \text{exp}\left(\ii\frac{2n\pi}{\cN}\Gamma_5\right)\lambda\,,\quad n=0,1,2,\ldots, \cN-1.
\end{equation}
Lattice simulations of four-dimensional $\sN=1$ SYM show that chiral symmetry is indeed 
spontaneously broken at zero temperature and restored above a critical temperature 
\cite{Bergner:2014saa}. 

The two-dimensional $\sN=(2,2)$ SYM theory can be derived from the 
four-dimensional theory via a Kaluza-Klein torus reduction. Thereby one
compactifies two directions on a torus such that $\mathbb{R}^4 \to \mathbb{R}^2 \times \mathcal{T}^2$ 
and assumes, that the fields are constant on the torus, e.g.
$\partial_M \lambda=0$ for $M=2,3$. The remaining non-compact coordinates
are $x^\mu$ with \mbox{$\mu\in \{0,1\}$}. Although the reduction does not 
depend on the particular representation of the four-dimensional 
$\Gamma$ matrices, it is  convenient to choose a particular one:
\begin{equation}
\Gamma_\mu=\mathbb{1} \otimes \gamma_{\mu}, \quad 
\Gamma_2=\ii\sigma_1 \otimes \gamma_5, 
\quad \Gamma_3=\ii\sigma_3\otimes\gamma_5, \quad 
\Gamma_5=\sigma_2\otimes \gamma_5
\label{Representation Gamma}
\end{equation}
with $\gamma_5=\gamma_0\gamma_1$.
In this representation, the charge conjugation matrices in two and four dimensions 
are related as $\cC_4=\mathbb{1}\otimes \cC_2$ and satisfy
\begin{equation}
\cC_2 \gamma_\mu \cC_2^{-1}=-\gamma_\mu^\trnsp\quad\Longrightarrow\quad
\cC_4 \Gamma_M \cC_4^{-1}=-\Gamma_M^\trnsp . \label{chiral symmetry}
\end{equation}
In a Majorana representation with purely real or imaginary $\gamma_\mu$ we
may choose $\cC_2=-\gamma^0$.
Applying the dimensional reduction to the Yang-Mills Lagrangian yields
\begin{equation}
-\frac{1}{4}F_{MN}F^{MN}
=-\frac{1}{4}F_{\mu\nu}F^{\mu\nu}
+\frac{1}{2}D_\mu \phi_m D^\mu \phi_m
+\frac{g^2}{4}\left[\phi_m,\phi_n\right]\left[\phi^m,\phi^n\right],\label{mink action}
\end{equation}
where the first term on the right hand side is the two-dimensional
Yang-Mills Lagrangian, the second term a kinetic term for the 
two adjoint scalar fields $\phi_m=A_{m+1}$ with $m\in\{1,2\}$
and the third term a quartic interaction potential for the scalar fields. 
The kinetic term for the four-dimensional Majorana 
fermion decomposes in a two-dimensional kinetic part and a Yukawa interaction
between the Majorana fermion $\lambda$ and the scalar fields $\phi_m$,
\begin{equation}
\bar{\lambda}\,\Gamma^M D_M \lambda=\bar{\lambda}\,\Gamma^\mu D_{\mu}\lambda-\ii\,g\bar{\lambda}\,\Gamma^{m+1}\left[\phi_m,\lambda\right].
\label{DimRed Kinetic Term}
\end{equation}
Note, that the four-component Majorana spinor $\lambda$ turns into two (real) Majorana 
spinors in two dimensions (in two dimensions an irreducible spinor has two components only).
Later we will merge them into one complex two-component Dirac spinor. After rescaling all fields $A,\lambda$ and $\phi$ according 
to $A \to g^{-1} A$ and absorbing afterwards the volume of the 
compactified torus in the gauge coupling $1/g^2 \to V_\mathcal{T}/g^2$, 
we obtain the action of the two-dimensional $\sN=(2,2)$ SYM theory
\begin{equation}
\begin{aligned}
	S=\frac{1}{2g^2}\int \D^2 x \tr\left\{-\frac{1}{2} F_{\mu\nu}F^{\mu\nu} 
	 +\ii\bar{\lambda} \Gamma^\mu D_\mu \lambda+D_\mu\phi_m D^\mu\phi_m \right.
	  \\ \left. + \bar{\lambda}\, \Gamma^{m+1}\left[\phi_m,\lambda\right] +
	   \frac{1}{2}\left[\phi_m,\phi_n\right]\left[\phi^m,\phi^n\right]\right\} 
	\label{two dimensional reducible action},
\end{aligned}
\end{equation}
the Euclidean version of which we use in our lattice simulations. 
In a next step we combine the four components of the Majorana 
spinor $\lambda$ in two components of an irreducible Dirac spinor in two dimensions
and rewrite the action in terms of Dirac fermions and complex scalars. 
Then the symmetries of the model are transparent and we can
easily compare with the $\mathcal{Q}$-exact formalism 
\cite{Catterall:2004np}. With the ansatz
\begin{equation}
	\lambda=\sum\limits_{r=1}^2 e_r \otimes \chi_r\quad\Longrightarrow\quad
	\bar{\lambda}=\sum_{r=1}^2 e^\trnsp_r\otimes\bar{\chi}_r\,,
\end{equation}
where $\{e_1,e_2\}$ is a Cartesian basis of $\mathbb{R}^2$, on
which $\Gamma^0$ in (\ref{Representation Gamma}) acts trivially,
and $\chi_r$ are irreducible Majorana spinors in two dimensions, we obtain
\begin{multline}
	S=\frac{1}{2g^2}\int \D^2 x \tr\bigg\{-\frac{1}{2} F_{\mu\nu}F^{\mu\nu}+D_\mu\phi_m D^\mu\phi_m+ \frac{1}{2}
	[\phi_m,\phi_n][\phi_m,\phi_n] \\
	  +\,\ii\bar{\chi}_r \gamma^\mu D_\mu \chi_r
	  - \bar{\chi}_r (\ii\sigma_1)^{rs}
	   \gamma_5[\phi_1,\chi_s] 
	-\bar{\chi}_r (\ii\sigma_3)^{rs} \gamma_5 [\phi_2,\chi_s]\bigg\}
	\label{two dimensional irreducible action}
\end{multline}
that contains two flavours $\chi_r$ of Majorana fermions and two real scalar fields.
Introducing the Dirac fermion $\psi$ and the complex scalar $\varphi$
according to
\begin{equation}
\psi=\frac{1}{\sqrt{2}}\left(\chi_1+\ii\gamma_5\chi_2\right)\,,\quad 
 \bar{\psi}=\frac{1}{\sqrt{2}}\left(\bar{\chi}_1+\ii\bar{\chi}_2\gamma_5\right)
,\quad\varphi=\phi_1+\ii\phi_2\,, \label{Psi definition}
\end{equation}
we end up with
\begin{align}
	S&=\frac{1}{g^{2}}\int \D^2 x \tr\left\{-\frac{1}{4} F_{\mu\nu}F^{\mu\nu}
	 + \frac{1}{2}(D_\mu\varphi)^\dagger (D^\mu\varphi)
	 -\frac{1}{8}\big[\varphi^\dagger,\varphi\big]^2
	 \right. \nonumber \\
	&\hskip28mm\left. +\,\ii\,\bar{\psi} \gamma^\mu D_\mu \psi
	-\, \bar{\psi}P_+\left[\varphi,\psi\right] 
	-\,\bar{\psi}P_-\big[\varphi^\dagger,\psi\big] 
 \right\}
 \label{two dimensional dirac action}
\end{align}
with chiral projection operators $P_\pm=\left(1\pm \gamma_5\right)/2$.
When proving this result one may use that for two Majorana
spinors $\chi_1,\chi_2$ the trace of $\bar\chi_1[\varphi,\chi_2]+\bar\chi_2[\varphi,\chi_1]$ vanishes.
Under dimensional reduction, the four-dimensional Lorentz transformations 
in SO$(1,3)$ turn into two-dimensional Lorentz transformations and 
flavour rotations for the scalar fields \mbox{($R$-symmetry)}, i.e. 
\begin{equation}
	\textrm{SO}(1,3)\to \textrm{SO}_L(1,1)\times \textrm{SO}_R(2)\, ,
\end{equation}
and correspondingly Spin$(1,3)$ turns into Spin$(1,1)$
and $R$-transformations of the two spinor fields, 
generated by $\Sigma_{23}=-\sigma_3\otimes\id/2$. This $R$-symmetry
acts on the real fields as
\begin{equation}
\begin{pmatrix} \phi_1 \\ \phi_2\end{pmatrix} \to R(2\alpha) \begin{pmatrix} \phi_1 \\ \phi_2\end{pmatrix}\,,\quad
	\begin{pmatrix} \chi_1 \\ \chi_2\end{pmatrix} \to R(-\alpha) \begin{pmatrix} \chi_1 \\ \chi_2\end{pmatrix},
\end{equation}
where $R(\alpha)$ is a rotation with angle $\alpha$. The complex fields
transform as
\begin{equation}
\varphi\to\exp(2\,\ii\,\alpha)\varphi\,,\quad
	\psi\to \exp(-\ii\,\alpha\,\gamma_5)\psi,\quad 
	\bar\psi\to \bar{\psi}\exp(-\ii\,\alpha\,\gamma_5)\,,
\end{equation}
which is identified as chiral symmetry in two dimensions. In contrast,
the four-dimensional chiral symmetry turns into 
a phase rotation of the Dirac field,
\begin{equation}
 	\lambda'=\exp(\ii\alpha\,\Gamma_5)\lambda=
 	\begin{pmatrix}\cos\alpha&\gamma_5\sin\alpha\\ 
 	-\gamma_5\sin\alpha&\cos\alpha
 	\end{pmatrix}
 	\begin{pmatrix}
 	\lambda_1\\ \lambda_2
 	\end{pmatrix}
 	 \quad \Rightarrow \quad \psi'=\exp(-\ii\,\alpha)\psi 
\end{equation}
and implies fermion number conservation in two dimensions.
This observation allows us to introduce two different fermion mass 
terms in the lattice formulation with Wilson fermions.
A four-dimensional Majorana mass term proportional to $\bar{\lambda}\lambda$ 
which violates fermion number conservation in two dimensions
or a two-dimensional Dirac mass term $\bar{\psi}\psi$ which violates chiral 
symmetry. When fine-tuning to the supersymmetric
continuum  limit we shall break chiral symmetry of the reducible model 
in order to have the same fermionic symmetries as in the 
$\mathcal{Q}$-exact formulation in \cite{Sugino:2003yb}, to which 
we shall compare our results.
\subsection{Expected mass spectrum}
\label{sec::ExpectedMassspectrum}
Veneziano and Yankielowicz were the first to derive a low energy 
effective Lagrangian for $\sN=1$ SYM theory in four dimensions,
in analogy to QCD \cite{Veneziano:1982ah}. They conjectured that
the lightest super-multiplet contains the bound states
shown in Table~\ref{tab::Multipletstructure4d}(a):
\begin{table}[htb]
\begin{subtable}[c]{0.5\textwidth}
\begin{tabular}{c|c|c}
	\toprule[1pt]
	particle & spin & name \\ \midrule[0.5pt]
	$\overline{\lambda}\gamma_5\lambda$ & 0  & a-$\eta$ \\
	$\overline{\lambda}\lambda$ & 0  & a-f \\
	$F_{MN}\Sigma^{MN}\lambda$ & $\frac{1}{2}$ & gluino-glueball \\ \bottomrule[1pt]
\end{tabular}
\subcaption{VY multiplet}
\end{subtable}
\begin{subtable}[c]{0.5\textwidth}
\begin{tabular}{c|c|c}
	\toprule[1pt]
	particle & spin & name \\ \midrule[0.5pt]
	$F^{MN}F_{MN}$ & 0 & $0^{++}$ glueball \\
	$F^{MN}\epsilon_{MNRS}F^{RS}$ & 0 & $0^{-+}$ glueball \\
	$F_{MN}\Gamma^{M}D^N\lambda$ & $\frac{1}{2}$ & gluino-glueball 
	\\ \bottomrule[1pt]
\end{tabular} 
\subcaption{FGS multiplet}
\end{subtable}
\caption{Multiplet structure of $\sN=1$ SYM theory as predicted by low energy effective actions \cite{Veneziano:1982ah,Farrar:1997fn}.}
\label{tab::Multipletstructure4d}
\end{table}
a scalar meson a-f, a pseudoscalar meson a-$\eta$ and a spin
$1/2$ bound state between a Majorana fermion and a gauge boson, called
gluino-glueball. We refer to this super-multiplet as the VY-multiplet. In a
confining theory one also expects glueballs in the 
particle spectrum. Therefore a second super-multiplet was added by Farrar,
Gabadadze and Schwetz \cite{Farrar:1997fn}. The FGS-multiplet is shown in Table~\ref{tab::Multipletstructure4d}(b). It contains a scalar glueball, a
pseudoscalar glueball as well as a spin $1/2$ gluino-glueball. 
Predictions about the mass-hierarchy of the two multiplets vary in 
the literature \cite{Veneziano:1982ah,Farrar:1998rm,Farrar:1997fn,Feo:2004mr}. 
In four dimensions large
scale Monte-Carlo simulations with Wilson fermions have been performed to
investigate the spectrum of bound states \cite{Bergner:2015adz}. The formation of
the VY-multiplet containing both mesons and a gluino-glueball has been observed
while the $0^{-+}$ glueball is significantly heavier. Within (large) errors the
$0^{++}$ glueball has the same mass as the f-meson, but due to mass mixing, it
is not clear whether the operator projects onto the correct state. Thus the
formation of a heavier multiplet has not been confirmed yet.

The multiplet structure of the $\sN=(2,2)$ SYM model can be 
extracted either from an effective Lagrangian of the two-dimensional system
or by dimensionally reducing the super-multiplets of the four-dimensional
effective theory. Thereby one should be cautious since the reduced
model should contain massless states \cite{Antonuccio:1998mq} and 
\begin{table}[h!]
\begin{center}
\begin{tabular}{c|c|c}
    \toprule[1pt]
	particle & spin & name \\ \midrule[0.5pt]
	$\overline{\lambda}\Gamma_5\lambda$ & 0  & a-$\eta$ \\
	$\overline{\lambda}\lambda$ & 0  & a-f \\
	$F_{\mu\nu}\Sigma^{\mu\nu}\lambda+2 \ii [\phi_1,\phi_2] \Sigma^{23}\lambda$ & $\frac{1}{2}$ & gluino-glue/scalarball \\\bottomrule[1pt]
\end{tabular}
\end{center}
\begin{center}
\begin{tabular}{c|c|c}
	\toprule[1pt]
	particle & spin & name \\  \midrule[0.5pt]
	$[\phi_1,\phi_2]F_{\mu\nu}$ & 0 & glue-scalarball  \\
	$F_{\mu\nu}F^{\mu\nu}-2D_\mu\phi_mD^\mu\phi_m-2[\phi_1,\phi_2]^2$ & 0 & $0^{++}$-glueball, scalarball\\
	$F_{\mu\nu}\Gamma^{\mu} D^\nu\lambda -D_\mu \phi_m \left(\ii \Gamma^{\mu}\left[\phi^m,\lambda\right]+ \Gamma^{m+1} D^\mu \lambda\right)$ & \multirow{2}{*}{$\frac{1}{2}$} & \multirow{2}{*}{gluino-glue/scalarball}\\
	$- [\phi_m,\phi_n]\Gamma^{m+1}\left[\phi^n,\lambda\right]$ & &  \\\bottomrule[1pt]
\end{tabular} 
\end{center}
\caption{Two-dimensional reduced super-multiplets for the $\sN=(2,2)$ theory. In the main body of the text we will call $F_{\mu\nu}\Sigma^{\mu\nu}\lambda$ the gluino-glueball and $[\phi_1,\phi_2] \Sigma^{23}\lambda$ the gluino-scalarball.}
\label{tab::reducemultiplets}
\end{table}
a super-multiplet with massless states looks different as a massive 
super-multiplet. Thus it is not straightforward to foresee the 
multiplet structure 
of the reduced system. In any case, the expected
bound states -- massive or massless -- of the $\sN=(2,2)$ SYM model are listed in Table~\ref{tab::reducemultiplets}.

\subsection{Supersymmetry restoration in the continuum limit}

As argued in the introduction, the lattice will break supersymmetry
explicitly. To restore it in the continuum limit, we have to fine-tune all
relevant supersymmetry breaking operators that are allowed by the remaining
symmetries on the lattice. For $\sN=(2,2)$ SYM, a discussion of
supersymmetry breaking operators is contained in \cite{Sugino:2003yb}. Thereby
the authors use a lattice formulation where one nilpotent
supersymmetry is exactly preserved on the lattice. In contrast, in 
our lattice formulation with Wilson fermions  the operator $\phi^2$ 
may show up in the effective action. To cancel this term
we must introduce a scalar mass counter-term $m_\text{s}^2\phi^2$ that 
has to be fine-tuned. The fine-tuned continuum value
$m_\text{s}^2=0.65948255(8)$ has been calculated to one-loop order
(which is sufficient for this theory) in \cite{Suzuki:2005dx}. 
Although a formulation with compact scalar
fields has been used, we checked that this value is also correct for 
non-compact scalar fields used in our simulation. 
This can be explained as follows:
The Jacobian of the transformation 
from the compact variables in \cite{Suzuki:2005dx} 
to non-compact variables cancels (in one-loop) the 
additional contribution in the action for the compact fields.
Thus we find the identical continuum value for $m_\text{s}^2$ in 
both formulations.

As for the four-dimensional mother-theory there is only one relevant 
susy  breaking term in two dimensions.  Because of the similarity of the two theories
one expects an important role of the fermion mass term
in two dimensions as well. Let us first recall the impact of a fermion
mass in four dimensions. Calculating the Ward identities for the 
chiral symmetry  and the supersymmetry on the lattice,
Curci and Veneziano demonstrated that only the renormalized gluino mass will 
appear as a relevant additional lattice contribution in the 
Ward identities \cite{Curci:1986sm}. Therefore by fine-tuning the bare 
gluino mass (in our case the fermion mass), one recovers
chiral symmetry and supersymmetry in the same limit. We expect
the same mechanism to be at work in two dimensions and thus
will fine-tune the fermion mass. Note that this idea is in line 
with \cite{Sugino:2003yb}, as the fermion mass must vanish in the 
continuum limit to recover the chiral limit, as it is not a relevant 
operator. A fine-tuning on the lattice will act as an improvement, 
reducing further supersymmetric violating contributions 
for finite lattice spacing.

\subsection{Euclidean formulation}

Since we can not simulate a model with Minkowski spacetime, we must
construct a continuation to the corresponding Euclidean theory. 
This continuation for theories with Majorana fermions was discussed in \cite{Montvay:2001ry,Nicolai:1978vc,vanNieuwenhuizen:1996tv}. In contrast 
to Dirac fermions there is only one Majorana spinor
with $\overline{\lambda}=\lambda^\trnsp\cC$. One cannot impose
the reality condition $\overline{\lambda}=\lambda^\dagger$. 
The action picks up an overall negative sign leading to
\begin{equation}
	S=\int \D^4x\, \mathcal{L},\qquad
	\mathcal{L}=\tr \left(\frac{1}{4}F_{MN}F^{MN}+\frac{1}{2}\bar{\lambda}\,
\Gamma^M D_M \lambda\right)\label{euclModel4}
\end{equation} 
with Euclidean Gamma-matrices $\Gamma_M$. Majorana fermions exist
in the dimensionally reduced Euclidean theory. As convenient
representation we may use 
\begin{equation}
\Gamma_\mu=\mathbb{1} \otimes \gamma_{\mu}, \quad 
\Gamma_2=\sigma_1 \otimes \gamma_5, 
\quad \Gamma_3=\sigma_3\otimes\gamma_5, \quad 
\Gamma_5=-\sigma_2\otimes \gamma_5\,,
\label{Representation Gamma2}
\end{equation}
now with Euclidean $\gamma_\mu$. The hermitean matrices $\Gamma_5=\Gamma_0\Gamma_1\Gamma_2\Gamma_3$ and 
$\gamma_5=\ii\gamma_0\gamma_1$ are related as 
$\Gamma_5=-\sigma_2\otimes \gamma_5$. 
Rescaling the fields and absorbing the volume of the torus
in a dimensionful gauge coupling the Lagrangian of the reduced
Euclidean model reduces to
\begin{equation}
\mathcal{L}
=\frac{1}{2g^2}\tr\left\{\frac{1}{2}F_{\mu\nu}^2
+\big(D_\mu \phi_m\big)^2
-\frac{1}{2}\left[\phi_m,\phi_n\right]^2
+\bar{\lambda}\Gamma^\mu D_{\mu}\lambda
-\ii\bar{\lambda}\,\Gamma^{m+1}\left[\phi_m,\lambda\right]\right\}\,.\label{euclaction}
\end{equation}
In terms of complex fields $\psi$ and $\varphi$ it takes the form
\begin{align}
	\mathcal{L}&=\frac{1}{g^2}\tr\left\{\frac{1}{4} F_{\mu\nu}^2
	 + \frac{1}{2}(D_\mu\varphi)^\dagger (D^\mu\varphi)
	 +\frac{1}{8}\big[\varphi^\dagger,\varphi\big]^2
	 \right. \nonumber\\
	&\left. \hskip20mm
	+\,\bar{\psi} \gamma^\mu D_\mu \psi
	+\,\ii\, \bar{\psi}P_+[\varphi,\psi] 
	+\,\ii\,\bar{\psi}P_-[\varphi^\dagger,\psi] 
 \right\}\,.
 \label{eucl2daction}
\end{align}
In actual simulations we choose the formulation (\ref{euclaction})
with two real scalar fields and a reducible four-component
Majorana spinor.

\subsection{Ward identities}
\label{sec::Wardidentities}
In order to check for the restoration of supersymmetry in the continuum limit, 
we monitor supersymmetric Ward identities
\begin{equation}
\langle \mathcal{QO}\rangle = 0,
\end{equation}
with supercharge $\mathcal{Q}$ introduced in \eqref{SUSY Trafo 4d} 
and operators $\mathcal{O}$. In four dimensions the fermionic operator 
\begin{equation}
\mathcal{O}_a(x)=\tr_\text{c}\left\{\overline{\lambda}_b(x)
{\left(\Gamma^{MN}\right)^b}_a F_{MN}(x)\right\}\label{op_ward}
\end{equation}
is frequently used and gives rise to a bosonic Ward identity 
\cite{Catterall:2009it}. On a finite lattice with lattice constant $a$
supersymmetry is violated and in terms of the rescaled dimensionless 
lattice fields the approximate Ward identity reads
\begin{align}
\frac{1}{N_t N_s}
\langle S_\text{B}\rangle&=
\langle \mathcal{L}_\text{B}\rangle
=\frac{1}{4}\big\langle \tr F^{MN}F_{MN}\big\rangle
=-\frac{3}{8}\frac{1}{2}\big\langle\tr \bar{\lambda}\slashed{D}\lambda\big\rangle
+O\big(\beta^{-1}\big)\nonumber\\
&=\frac{3}{2}\left(N_c^2-1\right)+O\big(\beta^{-1}\big)=\frac{9}{2}
+O\big(\beta^{-1}\big),\quad 
\frac{1}{\beta}=(ag)^2\,.\label{ward4d}
\end{align}
We made use of the fact that by translational invariance
expectation values of densities do not depend on the site $x$.
The identity relates the expectation values of the bosonic and
fermionic parts of the action, up to a
one-loop term of order $1/\beta$ which originates from the violation
of supersymmetry.
Note that in the on-shell formulation, one obtains the factor of $\frac{3}{8}$ 
instead of the factor $\frac{1}{2}$ in the off-shell formulation 
\cite{Catterall:2009it}.

In accordance with the dimensional reduction we decompose the operator
(\ref{op_ward}) into three terms: one with $\{M,N\}$ being $\{m,n\}$, one with
$\{\mu,\nu\}$ and finally one with $\{m,\mu\}$ or
$\{\mu,m\}$. The corresponding three (two-dimensional) Ward identities read
\begin{equation}
\begin{aligned}
	W_1=&\frac{1}{2}\big\langle\left[\phi_1,\phi_2\right]^2\big\rangle-
	\frac{\ii}{8}\big\langle\bar{\lambda}\Gamma_2\left[\phi_1,\lambda\right]+\bar{\lambda}\Gamma_3\left[\phi_2,\lambda\right]
	\big\rangle =0\,,\\
	W_2=&\frac{1}{4}\big\langle F_{\mu\nu}F^{\mu\nu}\big\rangle
	+\frac{\ii}{8}\big\langle\bar{\lambda}\,\Gamma_2\left[\phi_1,\lambda \right]-\bar{\lambda}\,\Gamma_3\left[\phi_2,\lambda\right]
	\big\rangle =\frac{3}{2},\\
	W_3=&\frac{1}{2}\big\langle D_\mu \phi^m D^\mu \phi_m\big\rangle=3.
\end{aligned} \label{Continuum Ward Identities}
\end{equation}

Note that the sum rule $W_1+W_2+W_3$ just reproduces the result $\frac{9}{2}$ in \eqref{ward4d}.

\section{Lattice formulation}
\label{sec::Lattice}

In the simulations we use Wilson fermions and the tree-level improved L\"uscher-Weisz gauge action \cite{Luscher:1984xn}. The scalar fields are treated as non-compact site-variables in the adjoint representation of the gauge group.
The action for the scalar fields is implemented by using the forward difference
\begin{equation}
	D^\text{f}_\mu \phi_x=\phi_{x+{e_\mu}}-U_{x,\mu}^\text{A}\phi_x
\end{equation}
in the kinetic term, where the link variables $U_{x,\mu}^\text{A}$ are in the adjoint representation. The fermion operator for Wilson fermions is
\begin{equation}
 D_{xy}=\left(m_\text{f}+2+\Gamma_{m+1}f^a \phi^m_a\right)\delta_{x,y}-\frac{1}{2}\sum \limits_\mu \left(\id-\Gamma_\mu\right)\delta_{x+e_\mu,y}U_{x,\mu}^\text{A}+\left(\id+\Gamma_\mu\right)\delta_{x-e_\mu,y}{U^\text{A}_{y,\mu}}^\trnsp
\end{equation}
where the matrices $(f^a)_{bc}$ are the structure constants of the gauge group $SU(2)$. Integration over the Majorana fermion yields the Pfaffian
of $\cC D$ and we obtain for the partition function as integral 
over the bosonic fields,
\begin{equation}
 Z=\int \mathcal{D}U\mathcal{D}\phi \,\Sign(\Pf(\cC D))\,\det(D^\dagger D)^{\frac{1}{4}} \,e^{-S[U,\phi]}.
\end{equation}
We made use of the $\Gamma_5$-hermiticity of the fermion operator $\Gamma_5 D \Gamma_5=D^\dagger$. The fourth root of $D^\dagger D$ is approximated by a rational approximation in the rHMC \cite{Kennedy:1998cu,Clark:2003na,Clark:2005sq,Clark:2006wq} algorithm.

\subsection{Sign problem and flat directions}
\label{sec::signproblem}
\begin{figure}[htb]
 \inputFig{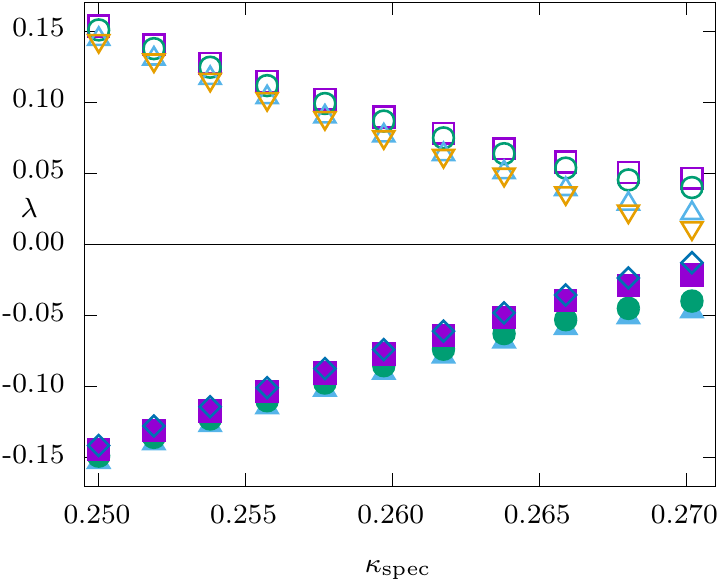}\hskip2mm
 \inputFig{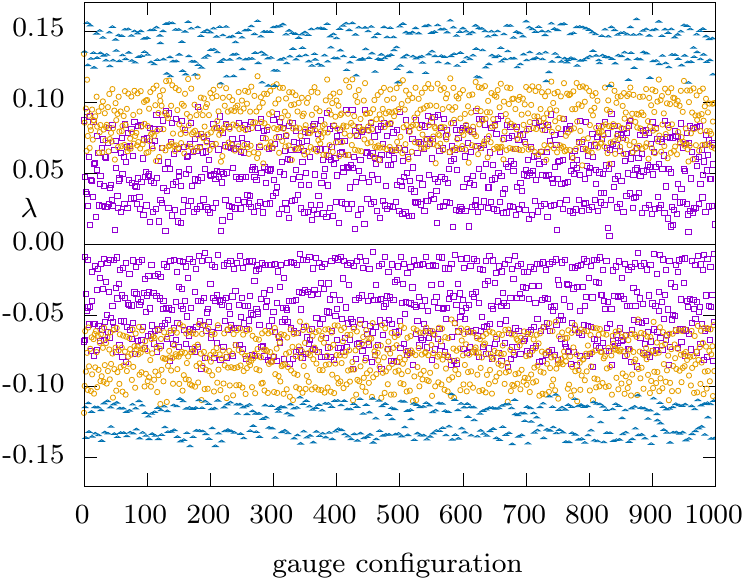}
\caption{Left: Spectral flow of $8$ eigenvalues with smallest 
absolute values for $\beta=15.5$, 
\mbox{$\kappa=0.27020$} on a $64\times 32$ lattice. 
Right: Smallest eigenvalues 
for three different values of the spectral flow parameter 
$\kappa_\text{spec}$:
$0.25379\,$ (blue triangles), 
$0.26174\,$ (orange circles) and
$\kappa\,$ (purple squares).
\label{Fig::SpectralFlow}}
\end{figure}

Two known problems may potentially spoil the Monte-Carlo simulations: 
a potential sign problem introduced by the Pfaffian and possible 
\emph{flat directions} in which the scalar potential is 
constant. We address both issues in turn.
Although the eigenvalues $\lambda_i$ of the hermitian matrix $Q=\Gamma_5 D$ 
are real and doubly degenerate \cite{Montvay:1997ak}, the Pfaffian can still 
introduce a sign problem that we have to take into account in the simulations. 
Using the dependence of the Pfaffian on the hopping parameter
$\kappa=1/(2\,m_\text{f}+4)$ it is possible to show \cite{Montvay:2001aj} 
that the Pfaffian and the determinant are related by 
\begin{equation}
\det D=\prod \limits_i \lambda_i^2 \quad \Rightarrow \quad 
\Pf(\cC D[U])=\prod\limits_i \lambda_i\,.
\end{equation}
We use the nice spectral flow method introduced in \cite{Montvay:2001aj} to 
monitor a potential sign problem. The idea is that for a given gauge 
field configuration (a typical one for fixed $\beta$ and $\kappa$)
the eigenvalues $\lambda_i$ vary continuously when the hopping parameter 
$\kappa_\text{spec}$ in the fermion operator increases. For the 
free operator with $\kappa_\text{spec}=0$
the Pfaffian is positive. Therefore, the Pfaffian can only become negative 
if an odd number of eigenvalues $\lambda_i(\kappa_\text{spec})$ 
change their signs as a function of $\kappa_\text{spec}$. We have 
monitored the $8$ eigenvalues with smallest absolute values, 
shown in the left panel of Figure~\ref{Fig::SpectralFlow} for 
configurations generated with $\beta=15.5$ and $\kappa=0.27020$ as 
function of the flow parameter $\kappa_\text{spec}$ 
increasing from $0$ to the value of interest $\kappa$.
The positive eigenvalues decrease monotonously while the negative 
eigenvalues increase as $\kappa_\text{spec} \to \kappa$, but they 
do not cross zero such that the Pfaffian for this configuration remains positive.
Furthermore we show the smallest eigenvalues for three ensembles of
$1000$ gauge configurations each belonging to the three flow parameters
$\kappa_\text{spec}=\kappa,\,0.25379,\,0.26174$ in Figure~\ref{Fig::SpectralFlow}. 
Even for $\kappa_\text{spec}=\kappa$ no eigenvalue is small
enough to change its sign. Hence the sign of the Pfaffian is always positive. 
We repeated the simulation for different volumes, inverse gauge couplings 
and hopping parameters. For $\kappa<\kappa_\text{c}$ we never 
observed a negative Pfaffian while for $\kappa>\kappa_\text{c}$ approximately 
one in thousand configurations had a negative sign. Thus we safely conclude 
that there is no sign problem in our simulations.
\\
The scalar potential 
\begin{equation}
 V[\phi_1,\phi_2]=\left[\phi_1,\phi_2\right]^2
\end{equation}
in the bosonic action is invariant under a shift
\begin{equation}
	\phi_1\to\phi_1+\alpha\,\phi_2 \qquad \phi_2\to\phi_2\,,
\end{equation}
where $\alpha$ is an arbitrary real parameter. This is an example of
a \emph{flat direction}  in the space of fields $(\phi_1,\phi_2)$ along 
which the potential is constant. Flat directions are generic
for SYM theories with extended susy and may destabilize Monte-Carlo simulations 
since the scalar fields may escape 
along these directions. Flat directions may either be lifted
dynamically by quantum corrections or explicitly by 
introducing a mass term $m_\text{s}^2 \phi^2$.  
Actually, as emphasized earlier, on the lattice we \emph{must} 
introduce a mass term with finite $m_\text{s}$
to find the correct supersymmetric continuum limit. This term
(which is needed even for $a\to 0$) lifts the flat directions explicitly.
\begin{figure}[htb]
\inputFig{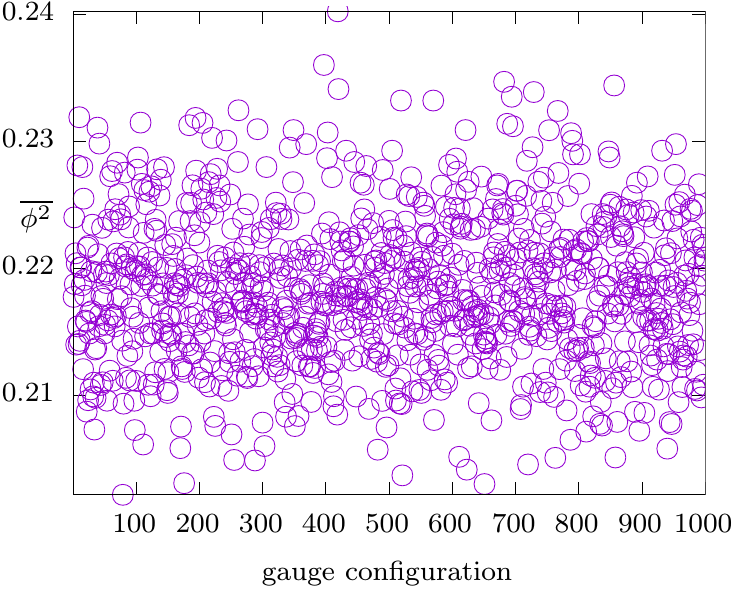}\hskip2mm
\inputFig{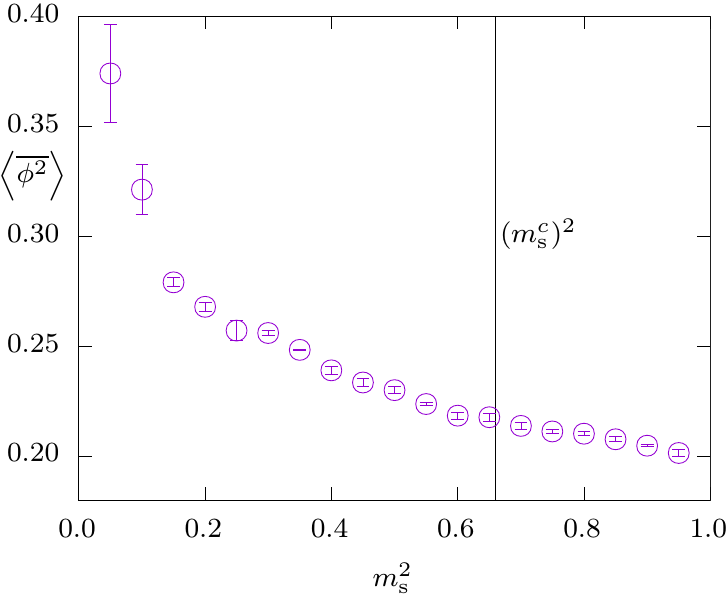}
\caption{Spatial average of squared scalar field as function of Monte-Carlo time for $\beta=14$, $\kappa=0.27233$, $m^2_\text{s}=0.6594826$ (left) and its expectation as function of $m^2_\text{s}$ (right) on a $16\times 16$ lattice.
\label{Fig::FlatDirections}}
\end{figure}
This is shown in Figure~\ref{Fig::FlatDirections} where we plotted
the spatial average $\overline{\phi^2}=\frac{1}{V}\sum \phi_x^2$ 
as function of Monte-Carlo time for $\beta=14$, $\kappa=0.27233$ on a $64\times 32$
lattice in the left panel and the expectation value of $\overline{\phi^2}$
as function of $m_\text{s}$ in the right panel. 
For all sets of parameters considered, the absolute value of the scalar 
fields does not run away. Hence we conclude, that flat directions are
lifted for values $m_\text{s}$ near the value of the supersymmetric
model and thus cause no problems in the simulations.
In a previous work the lifting of flat directions
has been observed even for the susy-breaking value $m^2_\text{s}=0$ and 
small values of the inverse gauge coupling \cite{August:2016orf}.

\subsection{Scalar and fermion mass fine tuning}
\label{sec::msmffinetuning}

The scalar mass is the only relevant coupling that 
has to be fine-tuned to restore super\-symmetry in the continuum limit
(in two dimensions the fermion mass needs not be fine-tuned). Its value 
in the thermodynamic and continuum limit  is analytically known 
from one-loop perturbation theory  $m^2_\text{s}=0.65948255(8)$ \cite{Suzuki:2005dx}.
On the finite $64\times32$ lattice the mass is shifted towards the smaller
value  $m^2_\text{s}=0.62849$. In order to investigate the dependence
of expectation values on $m^2_\text{s}$ we performed 
simulations for a larger range $m^2_\text{s} \in [0,1]$. 
Although the scalar mass breaks supersymmetry explicitly, it turns out that 
within the statistical uncertainties the Ward identities are independent of the 
scalar mass. Therefore we set the scalar mass to the continuum value $m_\text{s}^2=0.6594826$.

In contrast to four-dimensional $\sN=1$ SYM theory, a fine-tuning of the bare
fermion mass $m_\text{f}$ is not necessary to restore supersymmetry in the continuum limit. Nevertheless we shall enhance the chiral properties on the lattice by 
tuning $m_\text{f}$ to its critical value $m^\text{c}_\text{f}(L,\beta)$, that
depends on the inverse gauge  coupling $\beta$ but
depends little on the lattice size.
In the continuum limit, the critical fermion mass should approach
$m^\text{c}_\text{f}=0$, in agreement with the results in
\cite{Suzuki:2005dx,Sugino:2003yb}. There are two straightforward
methods to determine the critical fermion mass on a finite lattice.
The first uses the order parameter for chiral symmetry $\left<\bar{\lambda}\lambda\right>$ and
defines $m^\text{c}_\text{f}$ by the peak position of the 
chiral susceptibility. The second method comes from an
analogy to QCD which is also employed in the four-dimensional $\mathcal{N}=1$
SYM theory \cite{Veneziano:1982ah,Donini:1997gf,Munster:2014cja}:
Although the pion is not a physical particle in the theory, one can define
its correlation function in a partially quenched setup (for details see also 
appendix \ref{Appendix_B})
which mimics a second Majorana flavour in $\mathcal{N}=1$
SYM. The pion mass is related to the renormalized gluino mass by
\begin{equation}
	m_q\propto m_\pi^2.
\end{equation}
We expect this relation to hold in two dimensions as well and define the critical
fermion mass at the value where the gluino mass vanishes. The results for the two
methods are given in Table~\ref{tab::criticalmass}. 
\begin{table}[htb]
\begin{center}
\begin{tabular}{c|ccccc}
\toprule[1pt]
$\beta$ & $14.0$ & $15.5$ & $17.0$ & $40$ \\ \midrule[0.5pt]
$m_\text{f}^\text{c}(\chi_s)$ & $-0.1738(8)$ & $-0.1595(7)$ & $-0.1488(4)$ & $-0.0757(4)$   \\
$m_\text{f}^\text{c}(\pi)$ & $-0.1730(11)$ & $-0.1615(6)$ & $-0.1511(7)$ & $-0.0756(7)$  \\\midrule[0.5pt]
$\beta$ & $60$ & $80$ & $100$\\ \midrule[0.5pt]
$m_\text{f}^\text{c}(\chi_s)$ & $-0.0553(3)$ & $-0.0448(3)$ & $-0.0380(5)$  \\
$m_\text{f}^\text{c}(\pi)$ & $-0.0542(4)$ & $-0.0433(26)$ & $-0.0365(6)$  \\\bottomrule[1pt]
\end{tabular}
\caption{Critical fermion mass $m_\text{f}^\text{c}$ for different $\beta$. To determine the
mass we use the chiral susceptibility and the mass of the pion ground state.}
\label{tab::criticalmass}
\end{center}
\end{table}
Both methods yield comparable values for the critical fermion 
mass. One observes that the fermion mass approaches the expected continuum value
from below.\newline
In the following section we show that $\sqrt{\beta}\propto a$. Therefore we extrapolate our results to the continuum with the ansatz
\begin{equation}
	m_\text{f}^\text{c}(\beta)=m_\infty+c_1\beta^{-e_1}+c_2\beta^{-e_2}\,. \label{fitcriticalmass}
\end{equation}
The coefficients $c_i$ encode lattice artifacts and in the continuum
limit $m_\text{f}^\text{c}(\beta\to \infty) =m_\infty$. Since $m_\text{f}^\text{c}(\beta)$ does not depend significantly on the lattice size, we also include simulations at $\beta=40,60,80,100$ on smaller lattices into the extrapolation. The results of the fits are shown in Table~\ref{tab::contcriticalmass}. We give two different values $\chi^2_\text{w}$ and $\chi^2$ for the goodness of the fit. The first $\chi^2_\text{w}$ was calculated including the errors for $m^\text{c}_\text{f}$ as weights in the fit and the second $\chi^2$ without weights. $\chi^2$ is much smaller, showing that the fit of the given ansatz to the data is very good, but the errors for the critical masses are probably underestimated\footnote{The errors given for the critical fermion masses $m_\text{f}^\text{c}$ include only fit errors but not statistical errors.}. Within uncertainties the values for $m_\infty$ are compatible with the expected result $m_\infty=0$.
\begin{table}[htb]
\begin{center}
\begin{tabular}{c|cccc|cc}
\toprule[1pt]
  $m_\infty$ & $c_1$ & $c_2$ & $e_1$ & $e_2$ & $\chi_\text{w}^2$ & $\chi^2$ \\ \midrule[0.5pt]
  $0.0051(26)$ & $-0.285(32)$ & $-1.44(8)$ & \underline{$1/2$} & \underline{$1$} & $1.33$ & $6.33\times 10^{-7}$\\
  $-0.0126(8)$ & $-2.64(5)$ & $5.48(69)$ & \underline{$1$} & \underline{$2$} & $2.31$ & $7.88\times 10^{-7}$\\
  $-0.0041(18)$ & $-1.48(6)$ & \underline{$0$} & $0.820(18)$ & - & $1.06$ & $5.42\times 10^{-7}$\\
\bottomrule[1pt]
\end{tabular}
\caption{Fit values for the fit function given in \eqref{fitcriticalmass}, for three different sets of parameters. The mass $m_\infty$ represents the continuum value of the critical fermion mass $m^c_\text{f}$, which should be zero. The underlined parameters are prescribed in the 2-parameter fits.}
\label{tab::contcriticalmass}
\end{center}
\end{table}

\subsection{Wilson loops and confinement}
\label{sec::WilsonLoops}
In order to determine the lattice spacing and perform the continuum limit in section~\ref{sec::lattice spacing}, we consider the static quark-antiquark potential
\begin{equation}
	V(r)=A+\sigma r
	\label{eq:potential}
\end{equation}
in the fundamental representation of SU(2) with 
the string tension $\sigma$.
In two dimensions the Coulomb potential is a linear function in $r$ and the L\"uscher term is absent.
Hence we do not expect a $1/r$ term for small and large separations of the static charges. For large separations of the charges the
potential may flatten if there is string breaking.
If there is screening by massless particles then
the string tension should vanish.

In this subsection we calculate the static
quark-antiquark potential to see whether the theory is confining or whether the fermions can screen the
external charges. In addition to the theory of
investigation we calculate the potential
for the simpler $\mathcal{N}=(1,1)$ SYM theory
in two dimensions.
The latter is obtained by a dimensional reduction of the three-dimensional $\mathcal{N}=1$ SYM theory
and its action in the continuum reads
\begin{equation}
	S=\frac{1}{2g^2}\int\text{d}^2x\,\tr\,\left\lbrace-\frac{1}{2}F_{\mu\nu}F^{\mu\nu}+\text{i}\bar{\lambda}\gamma^\mu D_\mu\lambda+D_\mu\phi D^\mu\phi - \text{i}\bar{\lambda}\gamma_5[\phi,\lambda]\right\rbrace\,.
\end{equation}
It contains one adjoint scalar $\phi$ as well as one adjoint Majorana fermion $\lambda$. The $\gamma^\mu$ are two-dimensional matrices
as they are for the three-dimensional mother theory.
It has been argued in \cite{Gross:1995bp,Armoni:1998kv} 
that in $\mathcal{N}=(1,1)$ SYM theory a cloud of 
massless gluinos screens a static quark in the fundamental
representation. When the gluinos become massive, supersymmetry 
is broken, screening disappears and confinement should
be observed. It is believed that this is
a generic feature of two-dimensional YM-theories with 
massless adjoint fermions.

\begin{figure}[htb]
		\inputFig{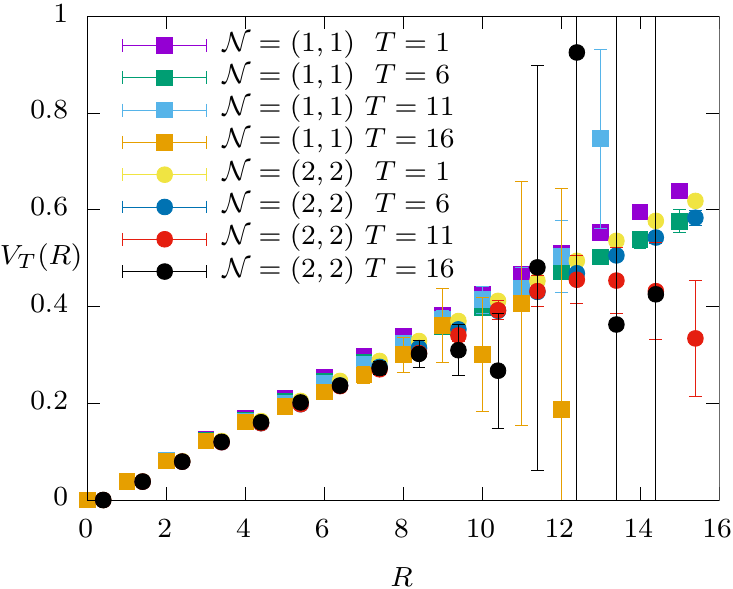}\hskip2mm
		\inputFig{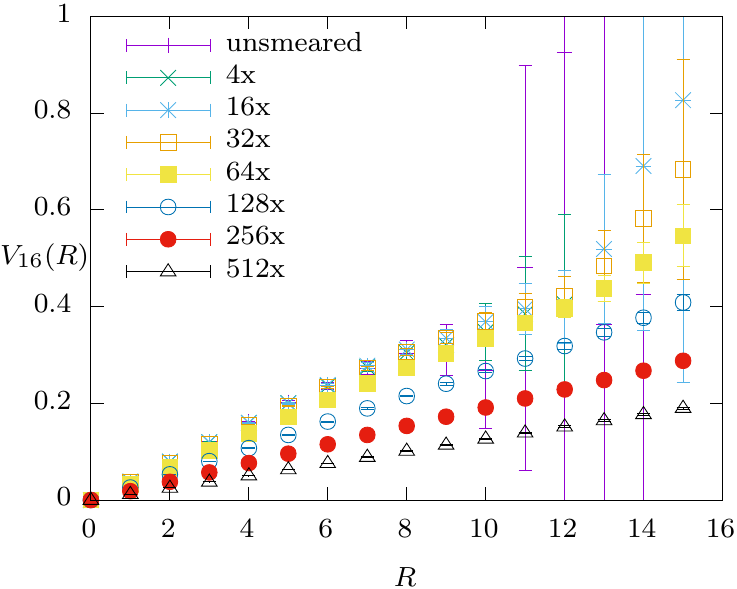}
		\caption{Left: Static fundamental quark-antiquark potential of \mbox{$\mathcal{N}\!=\!(1,1)$} and \mbox{$\mathcal{N}\!=\!(2,2)$} SYM theory. The measurements where done for several temporal extends $T$ on the $64\times 32$ lattice including reweighting of the Pfaffian. The \mbox{$\mathcal{N}\!=\!(2,2)$} data is shifted slightly for clarity of presentation. Right: Comparison of different levels of STOUT smearing with smearing parameter $\epsilon=0.4$ for the \mbox{$\mathcal{N}\!=\!(2,2)$ SYM with temporal Wilson loop size $T=16$.}}
		\label{fig:potential}
\end{figure}

Our lattice results for the static quark-antiquark potential $V_T(R)=\log\left(\frac{W_{R,T}}{W_{R,T+1}}\right)$ with Wilson loops $W_{R,T}$ are shown in Figure~\ref{fig:potential}.
For both theories\footnote{The $\mathcal{N}=(1,1)$ SYM theory suffers from a mild sign problem which can be treated 
with the help of a exact reweighing by measuring the Pfaffian.}
we find a linear raising potential.
To suppress statistical fluctuations, we used different numbers of STOUT smearing steps. With more smearing the potential becomes flatter, since fluctuations on scales of the final broken-string
state are suppressed.
We also measured Wilson-loops with unusual small $T$ to amplify the
signal to noise ratio. For $T\approx R>10$ the errors become
large since the signals are exponentially suppressed.
The unsmeared and smeared data (with controlled
statistical errors) both show no evidence of string breaking.
If there would be screening for massless fermions, 
then the string tension should tend to zero for light
fermions.
We performed simulations for several values of the fermion mass $m_\text{f}\in[-0.1640,0.0]$ and always obtained a 
linear rising potential. The string tension
decreases approximately $10\%$ towards the chiral limit.
Hence there seems to be no signal of screening in the 
simulations.
It may be that the Wilson loop has a poor overlap to the broken-string ground state, but this seems unlikely since its behavior does not change even close to the chiral limit.
Another explanation could be, that in a compact formulation
of gauge theories certain states are projected out of
the Hilbert space and screening cannot occur.
Of course, for an affirmative answer we would need a larger
set of operators and higher statistics or even better,
a method similar to the multi-level L\"uscher-Weisz
algorithm with exponential error reduction, as it
exists for pure gauge theories \cite{Luscher:2001up,Wellegehausen:2010ai}.

\subsection{Scale setting and lattice spacing}
\label{sec::lattice spacing}
\begin{figure}[htb]
\inputFig{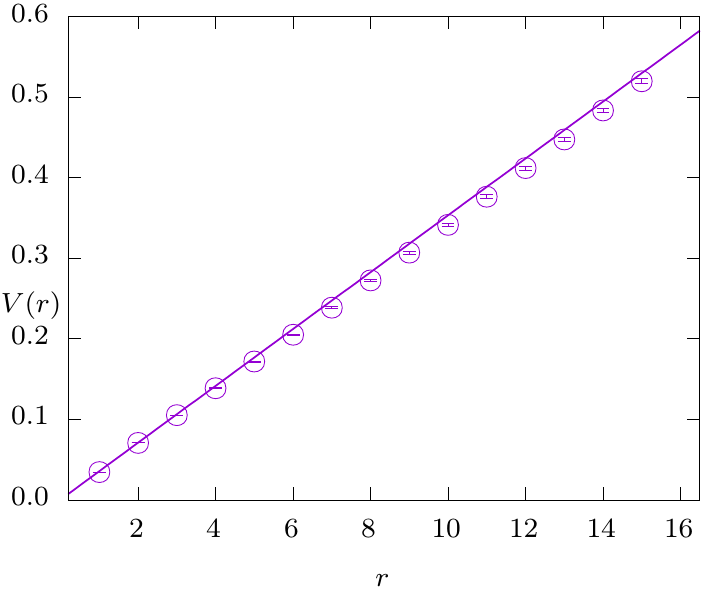}\hskip2mm
\inputFig{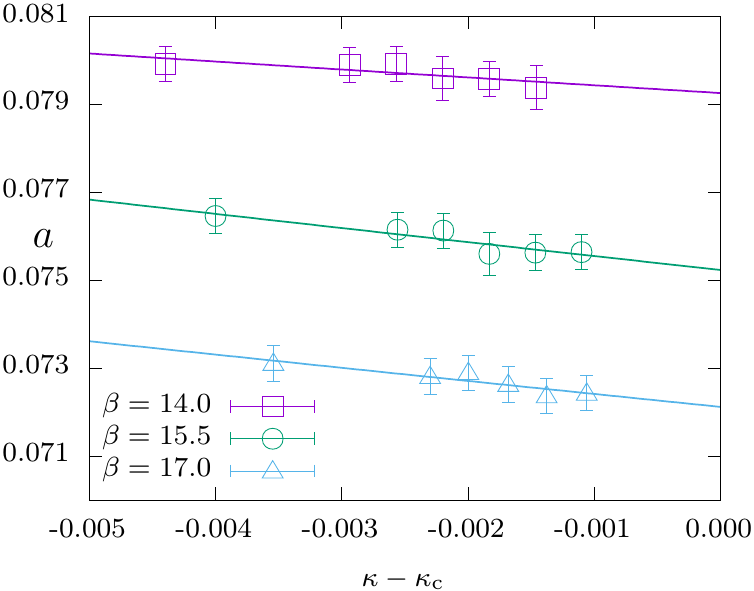}
\caption{Left: Static quark potential and fit to \eqref{eq:potential} for $\beta=17.0$ and $\kappa=0.26655$.
Right: Lattice spacing $a$ for $\beta=14.0,\,15.5$ and $17.0$ as function of $\kappa$
on a $64\times 32$ lattice.\label{Fig::ScalarPotential}}
\end{figure}

In order to determine the lattice spacing and perform the continuum limit, 
we consider the static quark-antiquark potential in the fundamental 
representation of SU(2) and extrapolate with the expected form \eqref{eq:potential} to the chiral limit.
For $\beta=17$ and $\kappa=0.26655$ the
potential is shown in Figure~\ref{Fig::ScalarPotential}. 
To compare our results to usual QCD lattice data, we employ the Sommer scale
\cite{Sommer:1993ce} and define a lattice spacing in physical units. 
The results for three different values of the inverse gauge coupling 
\begin{equation}
	\beta=\frac{1}{a^2 g^2} \label{latticecoupling}
\end{equation}
are depicted in Table~\ref{tab::latticespacing}.
Since the lattice spacing $a$ depends on the fermion mass, we 
extrapolate the latter to its chiral limit $m_\text{f}=m^\text{c}_\text{f}$. The
results are given in Table~\ref{tab::latticespacing}. In the last rows
we checked that the inverse dimensional coupling $1/g^ 2=\beta a^2$
in \eqref{latticecoupling} is almost independent of $\beta$, 
confirming that the continuum limit is reached for $\beta \to \infty$.

\begin{table}[htb]
\begin{center}
\begin{tabular}{@{}rrr|rrr@{}}
\toprule[1pt]
\multicolumn{3}{c}{$\beta=14.0$} & \multicolumn{3}{c}{$\beta=15.5$} \\ \midrule[0.5pt]
$\kappa-\kappa_\text{c}$ & $ a [\text{fm}]$ & $ \beta a^2[\text{fm}]$ & $\kappa-\kappa_\text{c}$ & $ a [\text{fm}]$ & $ \beta a^2[\text{fm}]$\\
$-0.00440$ & $0.07993(4)$ & $0.08944(9)$  & $-0.00400$ & $0.07646(4)$ & $0.09062(9)$\\ 
$-0.00294$ & $0.07989(4)$ & $0.08935(9)$  & $-0.00256$ & $0.07612(4)$ & $0.08981(9)$\\
$-0.00257$ & $0.07993(4)$ & $0.08944(9)$  & $-0.00220$ & $0.07613(4)$ & $0.08983(9)$\\
$-0.00220$ & $0.07959(5)$ & $0.08838(11)$ & $-0.00183$ & $0.07560(5)$ & $0.08859(12)$\\
$-0.00183$ & $0.07958(4)$ & $0.08833(9)$  & $-0.00167$ & $0.07563(4)$ & $0.08866(9)$\\
$-0.00146$ & $0.07938(5)$ & $0.08822(11)$ & $-0.00110$ & $0.07564(4)$ & $0.08868(9)$\\ \midrule[0.5pt]
$0$ & $0.07926(322)$ & $0.08795(51)$ & $0$ & $0.07524(310)$ & $0.08774(47)$\\ \midrule[0.5pt]
\multicolumn{6}{c}{$\beta=17.0$} \\ \midrule[0.5pt]
$\kappa-\kappa_\text{c}$ & $ a [\text{fm}]$ & $ \beta a^2[\text{fm}]$ & $\kappa-\kappa_\text{c}$ & $ a [\text{fm}]$ & $ \beta a^2[\text{fm}]$\\
$-0.00354$ & $0.07311(4)$ & $0.09087(10)$ & $-0.00168$ & $0.07263(4)$ & $0.08968(10)$
\\
$-0.00230$ & $0.07281(4)$ & $0.09012(10)$ & $-0.00138$ & $0.07237(4)$ & $0.08904(10)$\\
$-0.00200$ & $0.07290(4)$ & $0.09034(10)$ & $-0.00106$ & $0.07243(4)$ &  $0.08918(10)$\\ \midrule[0.5pt]
$0$ & $0.07212(266)$ & $0.08842(38)$\\
\bottomrule[1pt]
\end{tabular}
\caption{Lattice spacing for different combinations of $\beta$ and $m_\text{f}$. In the last rows of each $\beta$ section we give the extrapolations to the chiral limit.}
\label{tab::latticespacing}
\end{center}
\end{table}

\subsection{Smearing}

We use three different types of smearing. For the scalar
fields we utilize the low pass filter for functions. This smearing process is
defined as
\begin{equation}
	\tilde{\phi}^n(x)=\left(1+\epsilon\Delta\right)\tilde{\phi}^{n-1}(x)\quad \text{with} \quad \tilde{\phi}^{0}(x)=\phi(x),
\end{equation}
where $\phi(x)$ is the scalar field, $\tilde{\phi}^n(x)$ is the smeared field
and $\epsilon$ is the smearing parameter.
For gauge fields we use STOUT smearing \cite{Morningstar:2003gk} and for the
fermionic sinks and sources we apply Jacobi smearing
\cite{Gusken:1989ad,Allton:1993wc}.
In Table~\ref{tab::numberconfigurations} we give the number of configurations
generated for the given sets of parameters $\{\beta,m_\text{f},m_\text{s}\}$
on a $64\times 32$ lattice. 
A large number of configurations is needed to extract the masses of
the ground- and excited states of the f-meson.
This is due to large fluctuations of the two scalar fields 
entering the fermion operator via the Yukawa terms which
give rise to strong fluctuations in the fermion correlators.

\begin{table}[htb]
\begin{center}
\begin{subtable}[t]{.45\textwidth}
\begin{tabular}[t]{llll}
\toprule[1pt]
$\beta$ & $m_\text{f}$ & $m_\text{s}^2$ & \# C \\\midrule[0.5pt]
14.0 & -0.1440 & 0.6594826  & 10000 \\
14.0 & -0.1550 & 0.6594826  & 10000 \\
14.0 & -0.1565 & 0.6594826  & 10000 \\
14.0 & -0.1590 & 0.6594826  & 10000 \\
14.0 & -0.1615 & 0.6594826  & 10000 \\
14.0 & -0.1640 & 0.6594826  & 10000 \\
15.5 & -0.1320 & 0.6594826  & 10000 \\
15.5 & -0.1420 & 0.6594826  & 10000 \\
15.5 & -0.1445 & 0.6594826  & 10000 \\
\bottomrule[1pt]
\end{tabular}
\end{subtable}
\begin{subtable}[t]{.45\textwidth}
\begin{tabular}[t]{llll}
\toprule[1pt]
$\beta$ & $m_\text{f}$ & $m_\text{s}^2$ & \# C \\\midrule[0.5pt]
15.5 & -0.1470 & 0.6594826 & 10000 \\
15.5 & -0.1495 & 0.6594826  & 10000 \\
15.5 & -0.1520 & 0.6594826  & 10000 \\
17.0 & -0.1242 & 0.6594826  & 10000 \\
17.0 & -0.1329 & 0.6594826  & 10000 \\
17.0 & -0.1350 & 0.6594826  & 10000 \\
17.0 & -0.1372 & 0.6594826  & 10000 \\
17.0 & -0.1393 & 0.6594826  & 10000 \\
17.0 & -0.1415 & 0.6594826  & 10000 \\
\bottomrule[1pt]
\end{tabular}
\end{subtable}
\caption{Number of Configurations (\# C) for the given parameters
$\beta,\,m_\text{f}$ and $m_\text{s}$ on a $64\times 32$ lattice.
\label{tab::numberconfigurations}}
\end{center}
\end{table}


\section{Restoration of Ward identities}
\label{sec::Ward}
\noindent
The simple continuum Ward identities \eqref{Continuum Ward Identities} 
do not hold on the lattice since (in our formulation) there
are just no supersymmetries which leave the lattice action invariant.
But in the continuum limit we must recover these identities
if we take the finite additive renormalization of the parameter
$m_\text{s}^2$ into account.

Inspired by the treatment of four-dimensional models in
\cite{Curci:1986sm,Taniguchi:1999fc,Farchioni:2001wx,Luckmann:1997,Galla:1999}
we impose three rules to define the lattice transformations:
\begin{enumerate}[noitemsep,nosep]
\item They become the continuum susy transformations 
in the continuum limit.
\item They commute with the gauge transformations.
\item The transformation of the covariant derivative is the lattice equivalent 
of the continuum counterpart.
\end{enumerate}
These rules allow us to reduce the plethora of possible lattice transformations 
acting on the lattice fields $\{U_\mu(x),\lambda(x),\phi_m(x)\}$ to a small set.
We choose the transformations
\begin{equation}
\begin{aligned}
	&\bar{\mathcal{Q}}^\alpha U_\mu(x)=\frac{a}{2} U_\mu(x) {(\Gamma_\mu)^\alpha}_\beta\, \lambda^\beta(x+ae_\mu)\,,\quad
	\bar{\mathcal{Q}}^\alpha U_\mu^\dagger(x)=-\frac{a}{2}{(\Gamma_\mu)^\alpha}_\beta\, \lambda^\beta(x+ae_\mu) U_\mu^\dagger(x)\,, \\
	&\bar{\mathcal{Q}}^\alpha \lambda_\beta=0\,,\hskip6mm
	\bar{\mathcal{Q}}^\alpha\bar{\lambda}_\beta=-{\left(\Gamma_{\mu\nu}\right)^\alpha}_\beta G^{\mu\nu}\,,\hskip6mm
	\bar{\mathcal{Q}}^\alpha \phi_m=\frac{1}{2}({\Gamma_{m+1})^\alpha}_\beta\lambda^\beta, \label{eq:latticeSusyTrafo}
\end{aligned}
\end{equation}
where all fields but $U_\mu$ carry the canonical dimensions in
four dimensions and $a^2G^{\mu\nu}$ is the clover plaquette.
Since the lattice action is not invariant the continuum 
Ward identities are deformed to lattice identities
\begin{equation}
 \erw{\bar{Q}O}=\erw{O\, \bar{Q}S_\text{lat}}\,,\label{lat_wid1}
\end{equation}
where the transformation of the Lagrangian is given by
\begin{equation}
\bar{Q}^\alpha \mathcal{L}_\text{lat}
=\frac{\beta}{2}\left\{\partial_\mu s_\mu^\alpha 
- \left(m_\text{f}-m^\text{c}_\text{f}\right)\,\chi_\text{f}^\alpha 
+ \left(m^2_\text{s}-\left(m^\text{c}_\text{s}\right)^2\right)\,\chi_\text{s}^{\alpha}\right\}+\mathcal{O}(a)\label{lat_wid2}
\end{equation}
with dimensional quantities $\mathcal{L}_\text{lat}$ and $\beta$. 
After summing over all lattice sites the contribution of the 
supercurrent
$s^\alpha_\mu$
vanishes, up to terms of order $\mathcal{O}(a)$. In addition, the terms
$\chi^\alpha_\text{f}$ and $\chi^\alpha_\text{s}$ 
represent corrections introduced by a nonzero fermion mass $m_\text{f}$
and scalar mass $m_\text{s}$ away from their critical values.
These terms are suppressed after fine-tuning the masses. 
Details of the calculation are given in Appendix \ref{Appendix_A}. Finally we obtain the lattice Ward identities in the chiral limit $m_\text{f} \to m_\text{f}^\text{c}$
\begin{equation}
 \begin{aligned}
	W_\text{B}=&\beta V^{-1}\langle S_\text{B}\rangle
	+ m_\text{s}^2 \langle\tr\phi^2\rangle \to \frac{9}{2}\,, \quad
	W_3=\frac{\beta}{2} \langle \tr D_\mu\phi^a D^\mu\phi_a\rangle
	+  m_\text{s}^2 \langle\tr\phi^2\rangle \to 3\,, \\
	W_2=&\frac{\beta}{4}\langle \tr F_{\mu\nu}F^{\mu\nu}\rangle
	+\beta\langle\tr\bar{\lambda}\Upsilon\big\rangle \to\frac{3}{2}\,, \quad
	W_1=\frac{\beta}{2}\langle\tr\left[\phi_1,\phi_2\right]^2\rangle
	-\beta\langle\tr\bar{\lambda}\,\Upsilon\rangle 
	\to 0\,,\label{ward_dec}
 \end{aligned}
\end{equation}
where we used the abbreviation
\begin{equation}
\Upsilon=\frac{\ii}{8}\big(
\Gamma_2\left[\phi_1,\lambda\right]+\Gamma_3\left[\phi_2,\lambda\right]\big)\,.
\end{equation}

\subsection{Extrapolation to the chiral limit}
\label{sec::ExchiralLimit}
\noindent
\begin{figure}[htb]
\inputFig{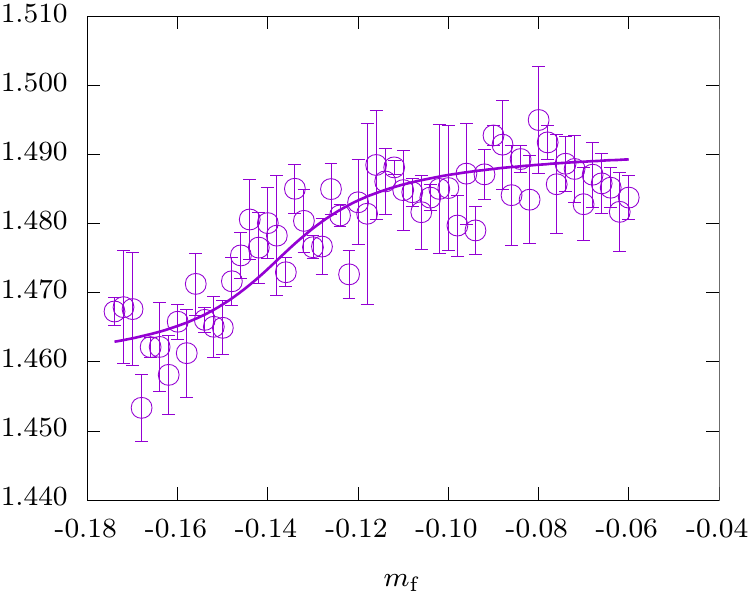}
\inputFig{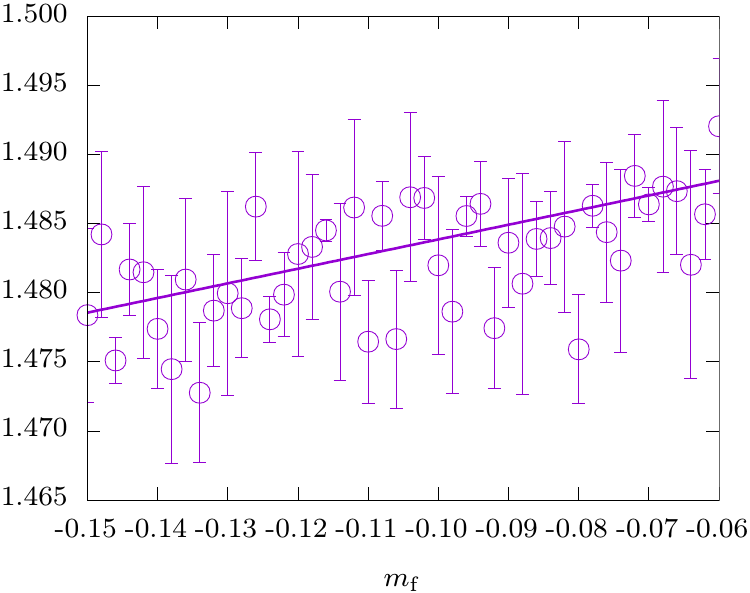}
\caption{The Ward identity $W_2$
in (\ref{ward_dec}) is shown for $\beta=17$ (left) and $\beta=40$ (right).}
\label{fig::W3fits}
\end{figure}

\noindent
We did check that 
the Ward identities show no dependence on the lattice size for $L_\text{S,T}>8$ for all $\beta$. Thus we simulated on
a moderate $32\times 16$ lattice
with parameters $\beta=40,60,80,100$.
To extrapolate our results to the chiral limit we need a guess
for the functional dependence of the Ward identities on the bare 
mass $m_\text{f}$. In two dimensions there is no spontaneous symmetry 
breaking and correlators are smooth functions of $m_\text{f}$.
Our simulations indicate that bosonic correlators show,
up to an additive constant $b$, a smoothed step function behavior
on the fermion mass. This motivates the following ansatz 
for their $m_\text{f}\,$-dependence near the 
\emph{critical} fermion bare mass $m_*$:
\begin{equation}
W(m_\text{f}) \sim a\,\arctan\left\{\xi\left(m_\text{f}-m_*\right)\right\}+b
\end{equation}
with fit parameters $a,b,m_*$ and $\xi$, where $\xi$ is to be interpreted 
as lattice correlation length. 

For example, in the left panel of Figure~\ref{fig::W3fits} we depicted
the arctan-fit to the Ward identity $W_2$ which is dominated by the term quadratic in the field strength tensor.
We observe that our ansatz yields a good approximation for the functional
dependence of the data on $m_\text{f}$. The extracted value 
for $m_*$ is very close to the critical fermion mass $m_\text{f}^\text{c}$. 
For $\beta\gtrapprox 40$ this ansatz is not appropriate anymore 
and we use a linear fit function, as seen on the right hand side 
of Figure~\ref{fig::W3fits}. These fits allow us to study
the Ward identities in the chiral limit.
\newline
Finally we have to extrapolate the Ward identities to the continuum limit. 
\begin{figure}[htb]
\inputFig{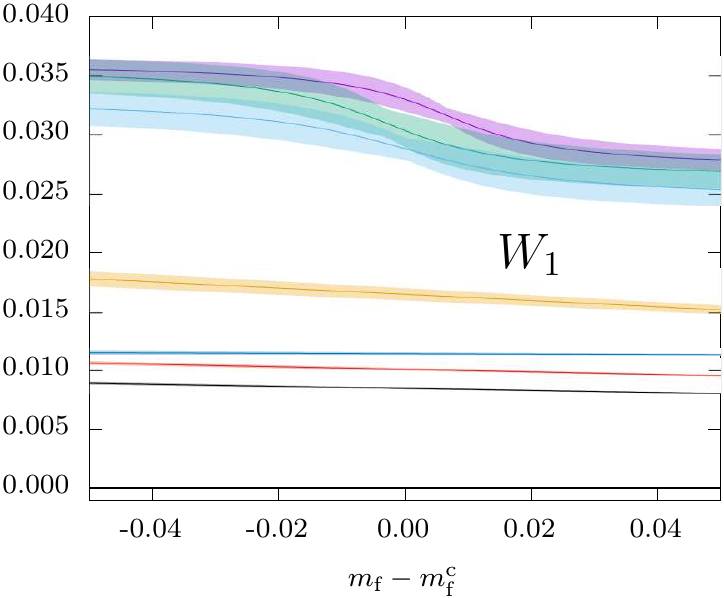}\hskip1mm
\inputFig{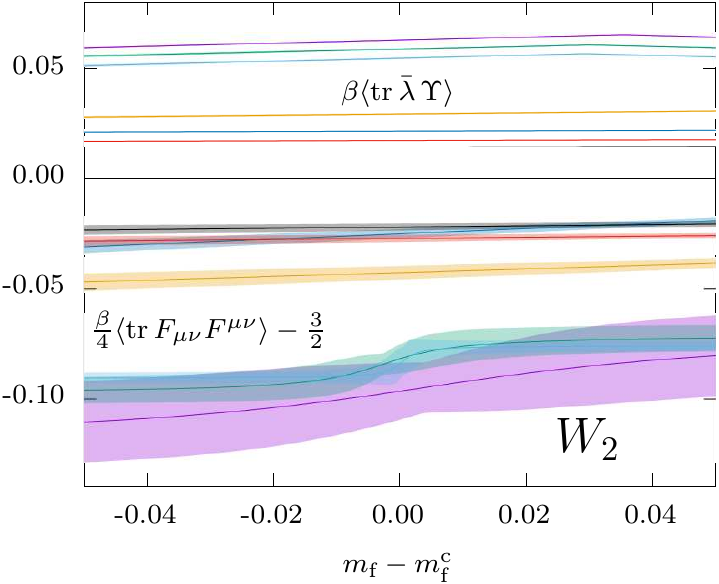}
\caption{Ward identities (\ref{ward_dec}) as functions of 
$m_\text{f}-m_\text{f}^\text{c}$ for 
various values of $\beta$ between $14$ and $100$. 
The colors represent different $\beta$: 14~\textcolor{b14}{$\bullet$}, 15.5~\textcolor{b155}{$\bullet$}, 17~\textcolor{b17}{$\bullet$}, 40~\textcolor{b40}{$\bullet$}, 60~\textcolor{b60}{$\bullet$}, 80~\textcolor{b80}{$\bullet$} and 100~\textcolor{b100}{$\bullet$}.
For $W_1$ (left panel) we show the fits and standard deviations (confident band). For $W_2$ (right panel) we show the two components $\beta\langle\tr\bar{\lambda}\Upsilon\big\rangle$ (upper half) and $\frac{\beta}{4}\langle \tr F_{\mu\nu}F^{\mu\nu}\rangle$ (lower half).}
\label{fig::FullWard}
\end{figure}
\begin{table}[htb]
\begin{center}
\begin{tabular}{c|cccc}
\toprule[1pt]
Ward identity & $W_1$ & $W_2$ & $W_3$ & $W_\text{B}$\\ \midrule[0.5pt]
$\beta=14.0$ & $0.0323(8)$ & $1.4678(79)$ &  $3.0222(5)$ & $4.5241(126)$ \\
$\beta=15.5$ & $0.0304(16)$ & $1.4732(118)$ & $3.0231(8)$ & $4.5298(143)$ \\
$\beta=17.0$ & $0.0288(10)$ & $1.4688(38)$ & $3.0185(9)$ & $4.5197(128)$ \\
$\beta=40.0$ & $0.0165(5)$ & $1.4834(6)$ & $3.0007(6)$ & $4.4867(11)$ \\
$\beta=60.0$ & $0.0123(1)$ & $1.4918(6)$ & $2.9968(8)$ & $4.5053(6)$ \\
$\beta=80.0$ & $0.0101(1)$ & $1.4901(6)$ & $2.9977(6)$ & $4.4973(9)$ \\
$\beta=100.0$ & $0.0085(1)$ & $1.4920(5)$ & $2.9972(6)$ & $4.5004(8)$ \\ \midrule[0.5pt]
$\beta\to\infty$ (Fit 1) & $-0.0053(3)$ & $1.5105(71)$ & $2.9773(66)$ & $4.4825(140)$ \\
$\beta\to\infty$ (Fit 2) & $0.0046(1)$ & $1.4981(46)$ & $2.9909(27)$ & $4.4936(74)$ \\
$\beta\to\infty$ (Fit 3) & $-0.0021(14)$ & $1.5507(872)$ & $3.0006(125)$ & $4.5492(1011)$ \\ \midrule[0.5pt]
$\beta\to\infty$ (weighted average) & $-0.0024(13)$ & $1.5267(424)$ & $2.9885(70)$ & $4.5128(507)$ \\ \midrule
theor. value & $0$ & $\frac{3}{2}$ & $3$ & $\frac{9}{2}$ \\ \bottomrule[1pt]
\end{tabular}
\caption{Values of Ward identities for different 
	values of $\beta$ on a $32\times 16$ lattice. The last five
	rows contain the continuum extrapolations with three different fit functions and a weighted average as well as the theoretical value \label{tab::Wardvalues} for unbroken susy.}
\end{center}
\end{table}

In Figure~\ref{fig::FullWard} we show the results for $W_1$ 
and the two contributions to $W_2$ in (\ref{ward_dec})
for different $\beta$.  In all cases we observe a monotonic convergence with increasing $\beta$.
\begin{figure}[!ht]
\begin{center}
\inputFig{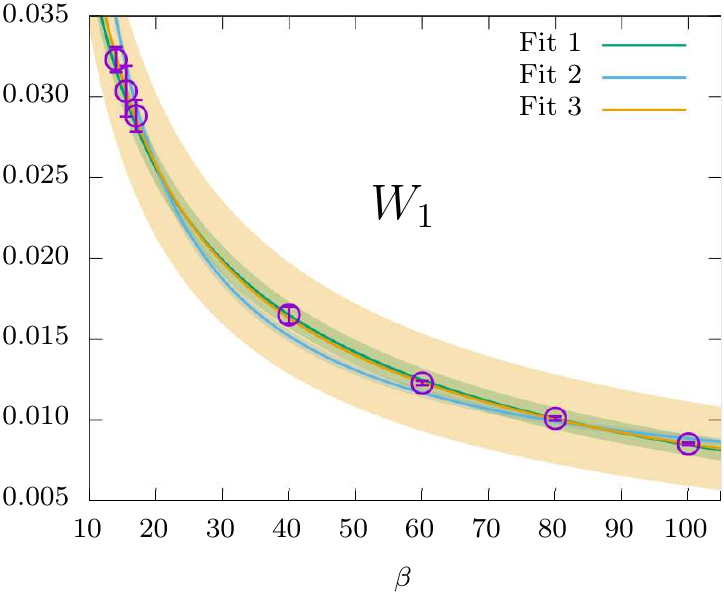}
\inputFig{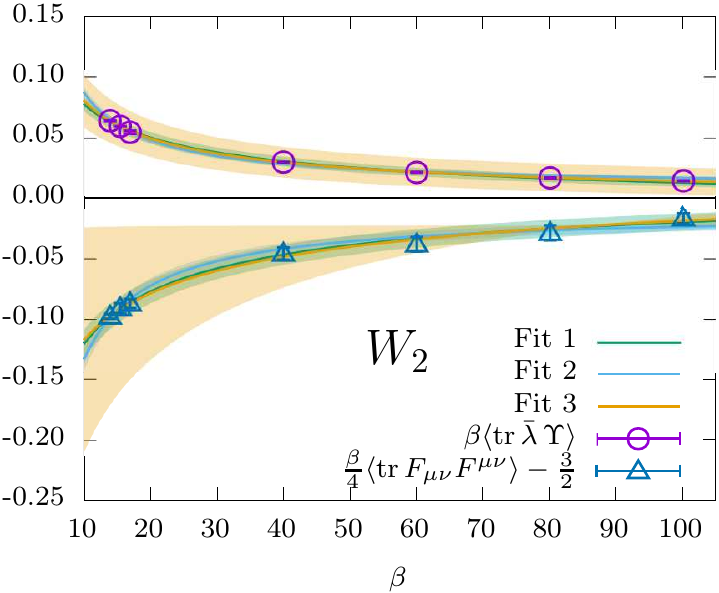}
\inputFig{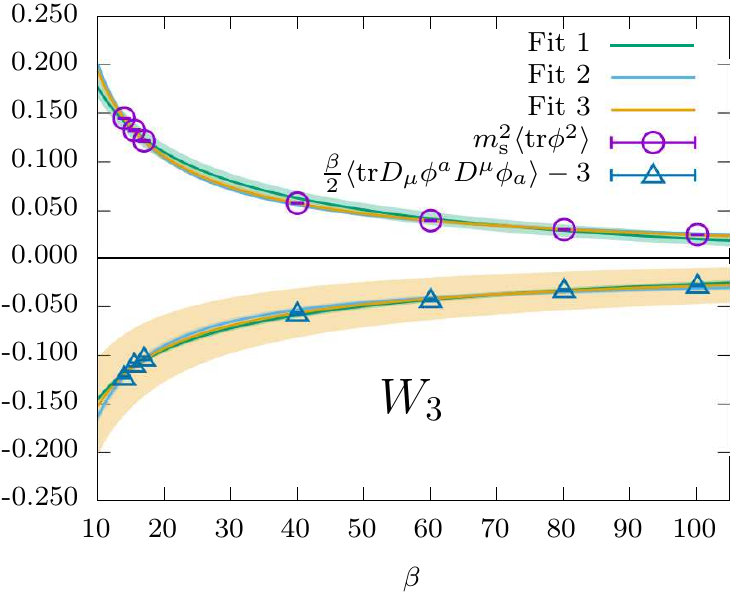}
\end{center}
\caption{Ward identity $W_1$ and various terms contributing
	to the identities $W_2$ and $W_3$ in (\ref{ward_dec})
	 for different values of $\beta$ together with three different fits used for the continuum extrapolation. The theoretical value in the supersymmetric continuum limit for $W_1$ is zero.}
\label{fig::Wvalues}
\end{figure}

In Table \ref{tab::Wardvalues} we listed the values of all Ward
identities for the chiral limit and different $\beta$ together with
the expected continuum value. The plots in Figure
\ref{fig::Wvalues} show the dependence on $\beta$. The Ward
identities clearly converge to the supersymmetric continuum value.
In order to extrapolate to the continuum limit, we use three different fits of the form
\begin{equation}
W(\beta)=W^\infty+b \, \beta^{-c}
\end{equation}
with the prescribed value $c=1/2$ for Fit 1 and $c=1$ for Fit 2 ($b$ and $W^\infty$ are free fit parameters). Fit 3 has three free fit parameters. The fits are shown in Figure~\ref{fig::Wvalues}. In Table~\ref{tab::Wardvalues} we give $W^\infty$ for $W_1$, the sum of the extrapolated components of $W_2$ and $W_3$ and the sum of these values for $W_\text{B}$.
From the three fit functions we can estimate a systematic error
coming from the choice of a particular fit function. 
This error alleviates our bias in choosing such a function. 
The \emph{weighted average} takes into consideration the goodness of 
the fits. The Ward identities clearly point to the restoration of supersymmetry in the continuum limit, indicating also no sign of spontaneous susy breaking.

\section{Mass spectrum}
\label{sec::Spectrum}
\noindent
In order to determine the mass spectrum of the theory, we first 
perform the infinite volume limit, then the chiral limit and finally the 
continuum limit.
For the infinite volume limit we study the dependence of
the mass of the lightest state on the size of the system
in order to locate a $\kappa$- and $\beta$-range where 
the results are (almost) insensitive to the volume.
Then we simulate the theory at a fixed lattice volume for different values of the hopping parameter $\kappa$ and extrapolate the results to the critical value $\kappa_\text{c}(\beta)$, where the gluino becomes massless. Finally we repeat the simulations for three different values of the gauge coupling $\beta$ and try to extrapolate the results to $\beta \to \infty$. 

\subsection{Volume dependence}

\begin{figure}[htb]
\inputFig{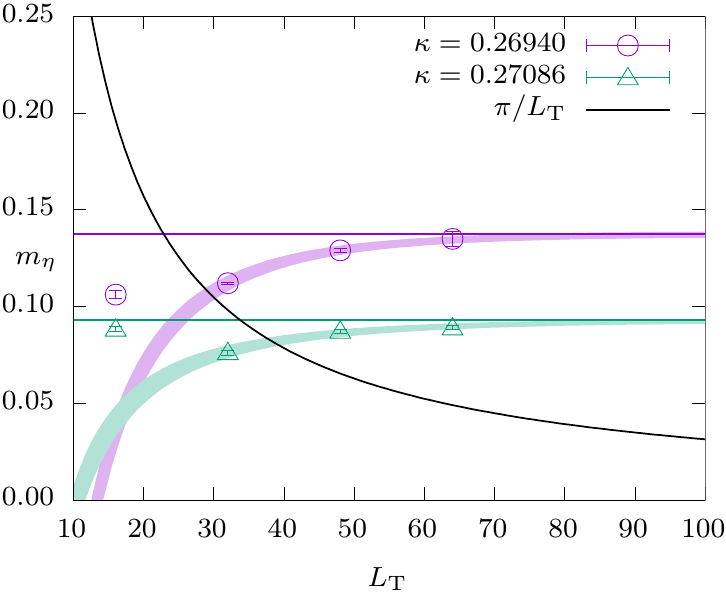}
\inputFig{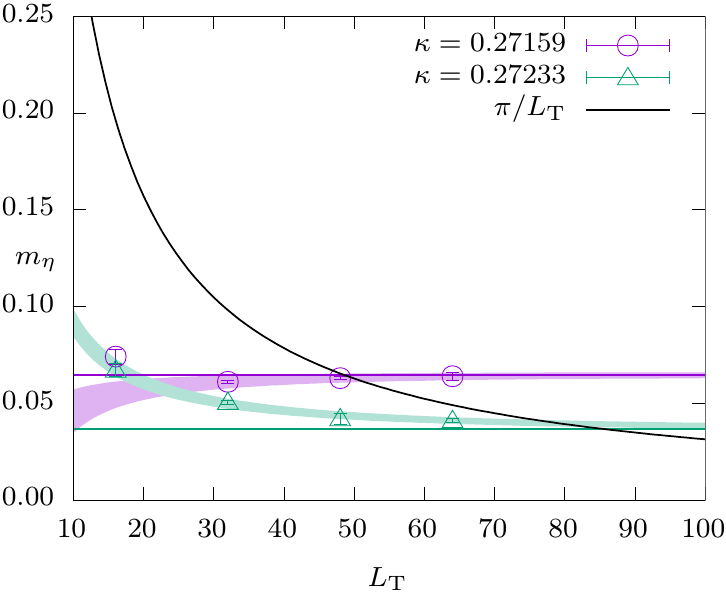}
\caption{Infinite volume extrapolation for the mass of the $\eta$-meson at $\beta=14$ and different values of the hopping parameter $\kappa$ compared to the smallest lattice momentum $\pi/L_\text{T}$. The horizontal lines indicate the infinite volume mass $m$.}
\label{fig:infinitevollimit}
\end{figure}

The finite volume dependency of bound states is given by \cite{Munster:1984zf,Luscher:1985dn}
\begin{equation}
	m_L=m-\frac{c}{L}\text{exp}\left(-\frac{L}{L_0}\right), \label{massextrapolation}
\end{equation}
where $m_L$ is the mass at a finite lattice with spatial length $L$ and $m$ the mass in the infinite volume limit. The parameter $L_0$ represents the scale at which finite volume effects set in. In order to eliminate this fit parameter, we 
relate it to the infinite volume mass of the lightest particle, i.e. 
$L_0=\pi/m_\eta$. The $\eta$-meson ground state mass $m_L$ is shown for $\beta=14$ and four different values of $\kappa$ in Figure~\ref{fig:infinitevollimit}.
We observe two different kinds of behaviour. For $\kappa=0.26940,0.27086$ and $0.27159$ the mass is monotonously increasing for $L_\text{T}\geq 32$. For $\kappa=0.27233$ it is monotonous decreasing. The explanation is that in the last case the infinite volume mass of $0.0365(14)$ is much smaller than the lattice cutoff $\pi/L_\text{T}$ for all lattices. If the mass gets close to the lattice cutoff, we get back the monotonously increasing function. Nevertheless, we observe that the fit function works well for all cases and yields reliable results for the infinite volume mass. For the largest $L_\text{T}$ value the mass $m_L$ is within statistical errors the same as the infinite volume mass $m$. Thus we will restrict ourselves to this lattice size for the spectroscopy.

\subsection{Mesons}

\begin{figure}[htb]
	\inputFig{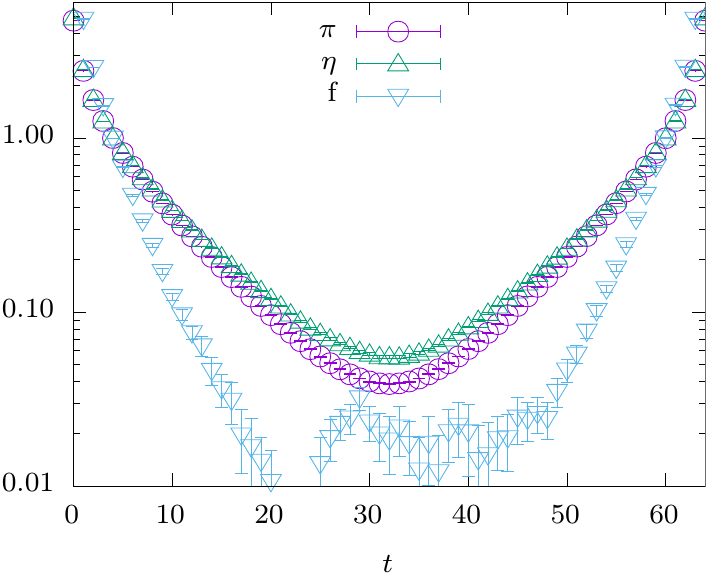}\hskip 5mm
	\inputFig{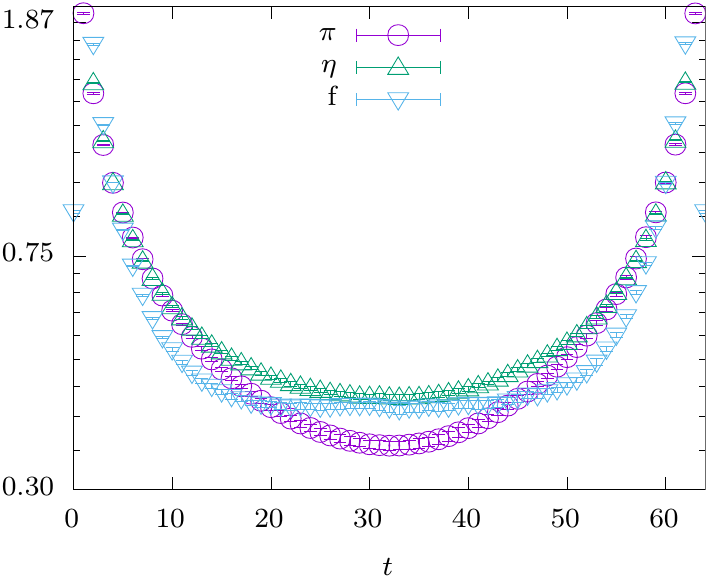}
	\caption{The $\eta\,$-,$\pi\,$- and f-meson correlation functions $C(t)$ as function of the temporal extend $t$ are shown for $\beta=17$ and $\kappa=0.26655$ (left) and $\kappa=0.26903$ (right).}
	\label{fig::chirallimit}
\end{figure}

We have calculated the  $\pi\,$-,\,$\eta\,$- and f-meson correlation function (see appendix \ref{Appendix_B}) for different values of the hopping parameter $\kappa$. In Figure~\ref{fig::chirallimit} we show
our results for two values of $\kappa\leq \kappa_\text{c}$. 
For the larger value $\kappa=0.26903$ the masses are slightly above the lattice momentum
cutoff. First of all we observe that the $\pi$- and the $\eta$-meson correlation
functions are very similar for all values of $\kappa$ considered 
and for intermediate values of $t$. For even larger $t$, the $\pi$ meson
correlation function decreases faster than the one for the $\eta$
meson. Thus the ground state of the latter must be lighter. As the
$\pi$ ground state mass becomes zero in the chiral limit, the same
will be true for the $\eta$-meson.

\begin{figure}[htb]
	\inputFig{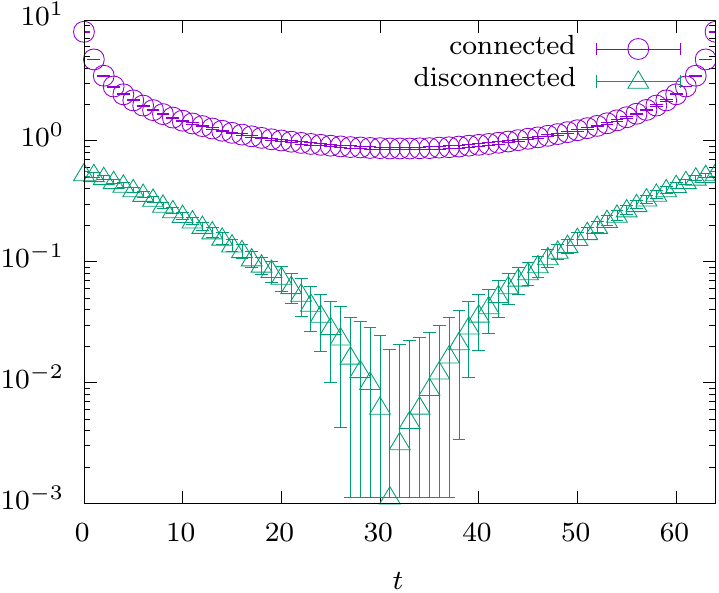}\hskip 5mm
	\inputFig{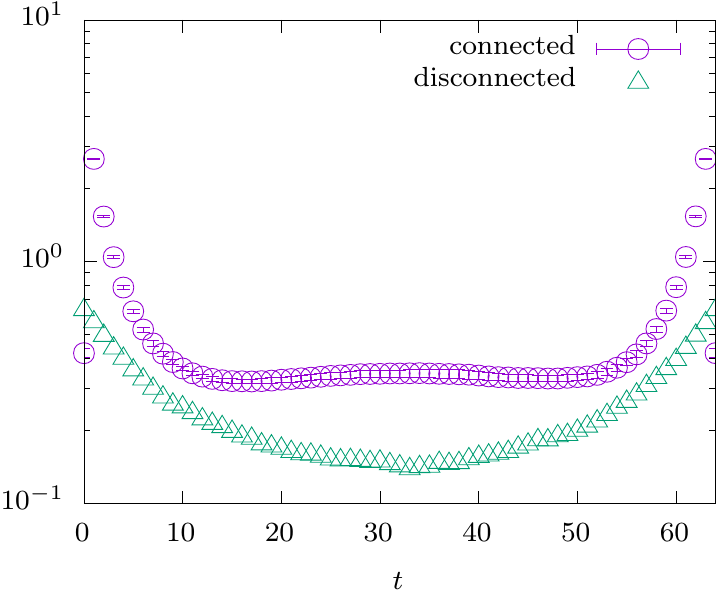}
	\caption{Connected and disconnected part of the $\eta$-meson (left) and f-meson (right) correlation function $C(t)$ as function of the temporal extend $t$ for $\beta=17$ and $\kappa=0.26903$.}
	\label{fig::partsofmesons}
\end{figure}

\begin{figure}[htb]
	\inputFig{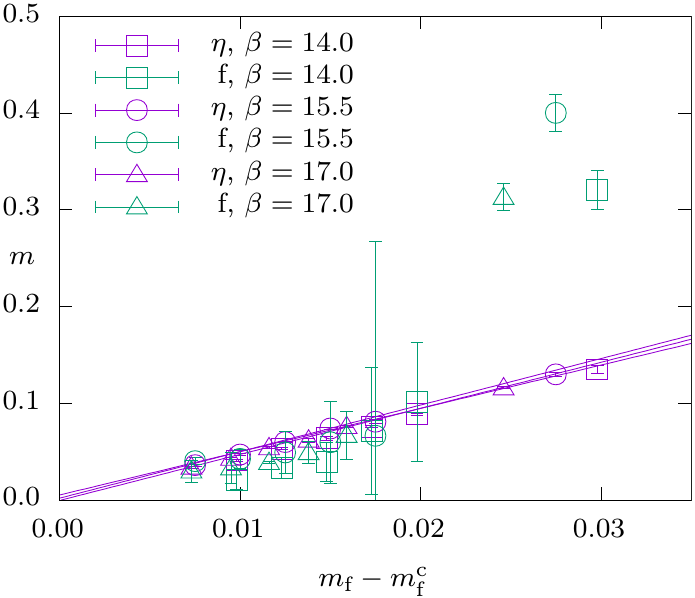}
	\inputFig{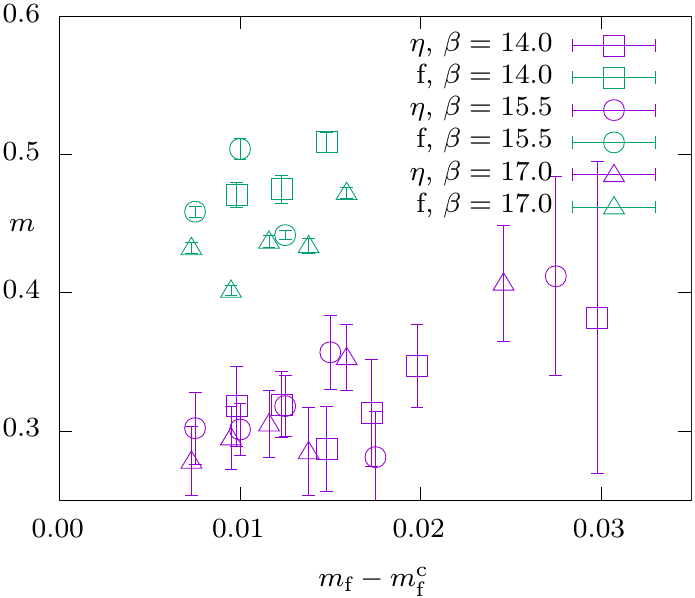}
	\caption{Ground (left) and excited (right) state masses of the $\eta$- and f-meson as function of the pion mass squared for $\beta=14$, $15.5$ and $17$. For the $\eta$ ground states we show linear fits.}
	 \label{fig::excitedmesons}
\end{figure}

Next we observe that the correlation functions for the f- and the $\eta$-meson become degenerate in the chiral limit. This suggests, that indeed both mesons form a multiplet in the chiral limit, independent of the restoration of susy in the continuum limit. To further investigate this behaviour we study
the connected and the disconnected contributions to the correlation functions.
Recall, that the pion correlation function is defined as the connected part of the $\eta$-meson correlation function.
In Figure~\ref{fig::partsofmesons} we depicted the two 
contributions to the correlation functions for the $\eta$-meson (left) and the f-meson (right).  
For the $\eta$-meson we find that the connected part is at least one order of magnitude larger than the disconnected part and thus
the $\eta$- and the $\pi$-meson correlation function are 
hard to distinguish. With increasing $t$ (but $2t\leq N_T$) 
the disconnected part gets even smaller. But despite
of this we can still
disentangle two slightly different masses in our simulations.
Only in the chiral limit will $\eta$ and $\pi$ both become massless.
For the f-meson the situation is different: the connected
and disconnected contributions are roughly of equal size
over the whole $t$ range. 
Hence a observed degeneracy between $\eta$-meson and f-meson 
correlation functions is nontrivial.
We determined the ground state and excited state masses of both mesons. The results are depicted in Table~\ref{tab::groundstates} 
as well as in Figure~\ref{fig::excitedmesons}.
\begin{table}[htb]
\begin{center}
\begin{tabular}{@{}l|lllllll@{}}
\toprule[1pt]
\multicolumn{5}{c}{$\beta=14.0$}  \\ \midrule[0.5pt]
$ \kappa $ & 0.26940 & 0.27086 & 0.27122 & 0.27159 & 0.27196 & 0.27233 \\  \midrule[0.5pt]
$ m_\eta $ & 0.135(4) & 0.089(1) & 0.076(1) & 0.064(2) & 0.053(1) & 0.041(1)  \\
$ m_\text{f} $ & 0.359(7) & 0.247(4) & 0.254(3) & 0.074(2) & 0.053(1) & 0.046(2) \\ 
$ m_{\eta^*} $ & 0.382(113) & 0.347(30) & 0.313(39) & 0.287(31) & 0.319(24) & 0.318(29)  \\
$ m_{\text{f}^*} $ & - & - & - & 0.509(7) & 0.475(10) & 0.471(9) \\
\midrule[0.5pt]
\multicolumn{5}{c}{$\beta=15.5$} \\ \midrule[0.5pt]
$ \kappa $ & 0.26767 & 0.26911 & 0.26947 & 0.26983 & 0.27020 & 0.27056 \\  \midrule[0.5pt]
$ m_\eta $ & 0.130(2) & 0.081(2) & 0.074(1) & 0.060(1) & 0.047(1) & 0.036(1) \\
$ m_\text{f} $ & 0.362(5) & 0.275(4) & 0.140(5) & 0.059(1) & 0.052(1) & 0.037(1)\\ 
$ m_{\eta^*} $ & 0.412(72) & 0.281(33) & 0.357(27) & 0.318(22) & 0.301(19) & 0.302(26) \\
$ m_{\text{f}^*} $ & - & - & 0.656(23) & 0.442(3) & 0.504(8) & 0.459(4) \\
\midrule[0.5pt]
\multicolumn{5}{c}{$\beta=17.0$} \\ \midrule[0.5pt]
$ \kappa $ & 0.26655 & 0.26779 & 0.26810 & 0.26841 & 0.26872 & 0.26903 \\  \midrule[0.5pt]
$ m_\eta $ & 0.116(1) & 0.076(1) & 0.062(2) & 0.054(1) & 0.043(1) & 0.034(2) \\
$ m_\text{f} $ & 0.335(2) & 0.094(2) & 0.064(3) & 0.052(4) & 0.030(1) & 0.034(1)\\ 
$ m_{\eta^*} $ & 0.407(42) & 0.353(24) & 0.285(32) & 0.305(24) & 0.295(23) & 0.278(25)\\
$ m_{\text{f}^*} $ & - & 0.473(4) & 0.434(5) & 0.437(4) & 0.402(4) & 0.433(4)\\
\bottomrule[1pt]
\end{tabular}
\caption{Masses of the $\eta$- and f-meson ground and excited states.}
\label{tab::groundstates}
\end{center}
\end{table}

We see that the mass of the $\eta$-meson ground state depends
linearly on the fermion mass $m_\text{f}$. In fact, the zero crossing
almost exactly hits the critical $m_\text{f}^\text{c}$. 
Thus $m_\eta$ is proportional to $m_\text{f}-m_\text{f}^\text{c}$
and will vanish in the chiral limit. For the $f$ meson, a
linear dependence is seen only for a fermion mass close to the critical fermion mass, where the latter is the same as for the 
pion. This behaviour is more pronounced for the larger 
values of $\beta$. Thus in the chiral limit we find the same 
ground state masses  for the f-meson and the $\eta$-meson.

For the excited states we can not make a comparably
strong statement, since it is more difficult to extract their masses.
For the $\eta$-meson we used a fit with three masses, which agrees
rather well over the whole $t$-range with the correlator. 
As the largest mass was above $1$, it is heavily 
afflicted with discretization artifacts and thus 
discarded. Hence only the masses of the ground states and 
first excited states are given in Table~\ref{tab::groundstates}.
We compared these results with the effective mass extracted
from the corresponding correlation function. For
both $\eta$ and $f$ we find one plateau corresponding
to their ground state mass. 
Using the so obtained values to fit the correlation
functions for small $t$ leads to the values of the first 
excited state of the f meson, given in Table~\ref{tab::groundstates}. 
Unfortunately this method of determination leads to a 
large unknown systematic error. Comparing the results for
different values of $\beta$, we observe that the mass of the excited f-meson decreases slowly with increasing $\beta$.
Thus it could approach the mass of the excited $\eta$ meson 
in the continuum limit. Unfortunately our results do not 
allow for an unambiguous  extrapolation to the chiral limit, preventing also the continuum extrapolation.

\subsection{Gluino-glueball}

In the four-dimensional multiplet we have two gluino-glueball particles, 
which differ by their transformation under parity. As interpolating fermionic  operator we use
\begin{equation}
	O_{GG}=\Sigma_{\mu\nu}F^{\mu\nu}\lambda
\end{equation}
where the $F^{\mu\nu}$ is approximated on the lattice by the clover plaquette.
Although the projectors on a definite parity quantum number 
are $P_\pm=(1\pm \Gamma_0)/2$ it is more convenient
to  project on periodic (S) and antiperiodic (A) correlation 
functions
\begin{equation}
	C_\text{A}(t)=\left< O_{GG}(t)O^\dagger_{GG}(0)\right>\,, \qquad C_\text{S}(t)=\left< O_{GG}(t)\Gamma_0 O^\dagger_{GG}(0)\right>.
\end{equation}
All other contractions over $\Gamma$-matrices can be written as a linear combination of these two correlation functions, as expected for two independent physical states.
\begin{table}[htb]
\begin{center}
\begin{tabular}{ccccccc}
\toprule[1pt]
$S$ & 12 & 40 & 120 & 200 & 300 & 400 \\ \midrule[0.5pt]
$m_\text{A}$ & 0.486(11) & 0.360(7) & 0.320(5) & 0.310(5) & 0.289(12) & 0.287(10) \\
$m_\text{S}$ & 0.410(10) & 0.313(5) & 0.265(2) & 0.252(3) & 0.246(3) & 0.243(3) \\
 \bottomrule[1pt]
\end{tabular}
\caption{Extracted masses for different smearing levels $S$ for the symmetric and antisymmetric gluino-glueball states for $\beta=17$ and $m_\text{f}=-0.1415$.}
\label{tab::smearinggg}
\end{center}
\end{table}
\begin{figure}
\center
\inputFig{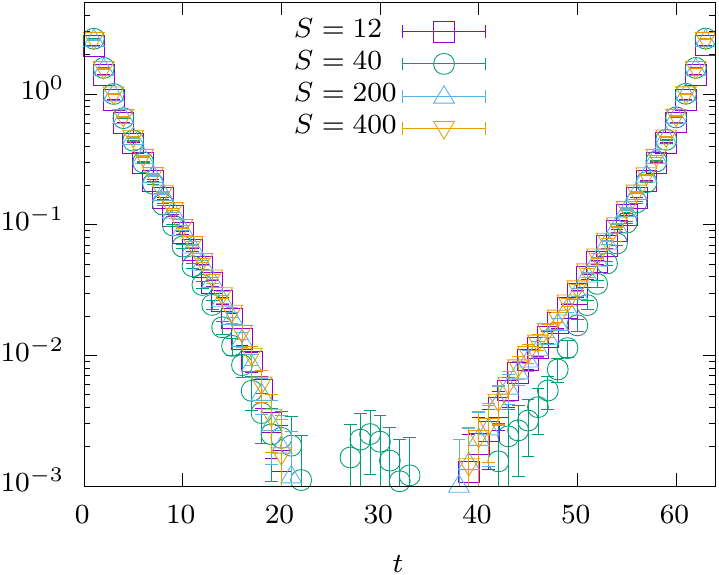}
\inputFig{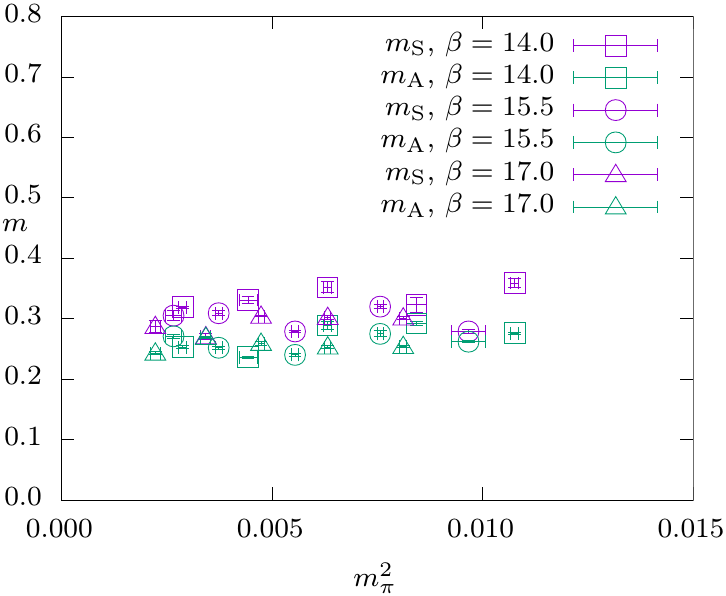}
\caption{Left: Gluino-glue correlation function $C(t)$ as function of the temporal extend $t$ at $\beta=17$ and $m_\text{f}=-0.1350$ for different smearing levels $S$. Right: Gluino-glue mass as a function of the squared pion mass.\label{fig::gluinogluesm}}
\end{figure}

The determination of masses on larger lattices is only possible with 
the help of gauge field smearing. We introduce the smearing level 
$S=\text{steps} \times \text{parameter}$, where 'steps' are the 
amount of smearing steps and 'parameter' is the smearing parameter 
for these steps. The correlation functions $C_\text{S}(t)$ for 
different smearing levels are shown in Figure~\ref{fig::gluinogluesm} 
(left panel). Even for a large number of smearing steps the signal 
still improves. Table~\ref{tab::smearinggg} shows our results for 
$\beta=17$ and $m_\text{f}=-0.1415$. For both masses, we see a 
nice convergence with increasing smearing. 
This behaviour is even seen for large smearing levels ($S=400$). 
Both masses $m_\text{A}$ and $m_\text{S}$ 
converge to the same value as expected in a parity symmetric theory. 
Furthermore  the mass depends only very weakly on the gauge coupling 
$\beta$ and the bare fermion mass $m_\text{f}$ (see 
Figure~\ref{fig::gluinogluesm}, right panel).

Comparing with the masses of the mesons, we find that the 
gluino-glueballs have comparable masses as the excited 
state of the $\eta$-meson.  An explanation for this 
unexpected behavior could be, that the first excited 
state of the gluino-glueball dominates the correlation function 
over a long $t$-range, such that the ground state contribution
is not visible on our lattice sizes. To see whether this is the case, we applied this large amount of smearing ($S=400)$, but we did not observe any sign of a lighter particle in this channel. Thus an alternative explanation could be, 
that we indeed detected the ground state of the gluino-glueball. 
But then one must explain why the gluino-glueball forms a multiplet with the excited 
mesons and not the mesons in their ground states. The fermionic state in the VY-multiplet is a mixture of the gluino-glue and a gluino-scalarball. Possibly the gluino-scalarball has a lighter mass.
Unfortunately, also with a large amount of smearing for the scalar field, we are not able to obtain an estimate for its mass.

\subsection{Glue- and scalarballs}

\begin{figure}[htb]
\center
\inputFig{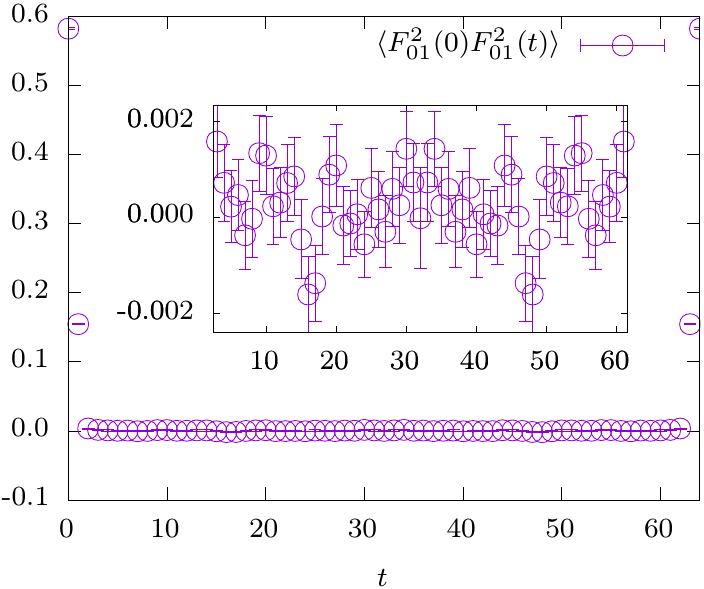}
\caption{Glueball correlation function $C(t)$ as function of the temporal extend $t$ for $\beta=17$ and $m_\text{f}=-0.074$.}
\label{fig::glueballcorr}
\end{figure}
The second multiplet of bound states consists of glue-, scalar- and  glue-scalarballs. 
The correlation functions of the corresponding 
interpolating operators show no correlation at all for large distances. For the glueball, this is shown in Figure~\ref{fig::glueballcorr}.
\begin{figure}[htb]
  \inputFig{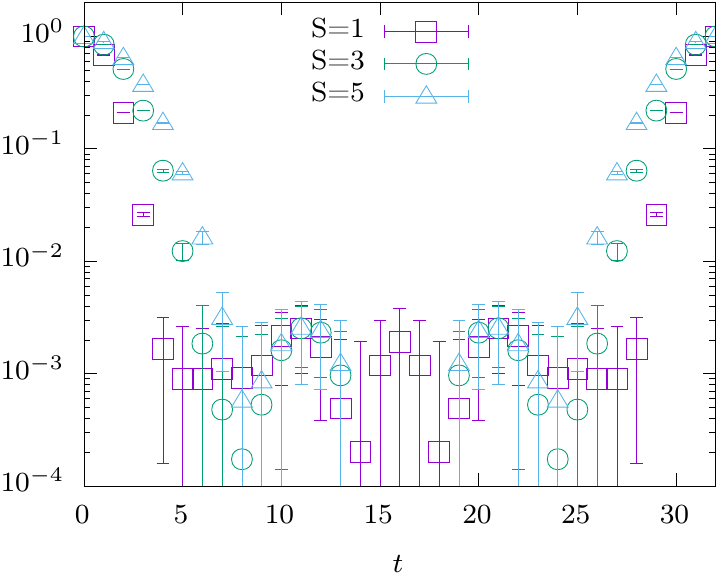}	
  \inputFig{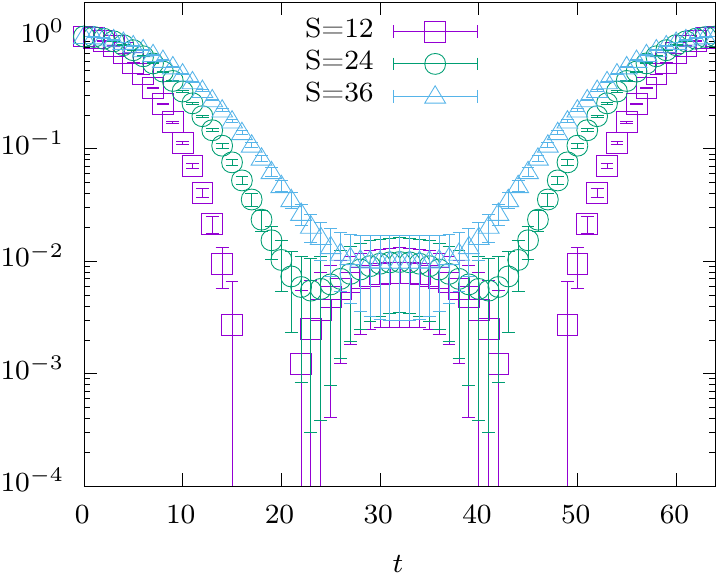}	
  \inputFig{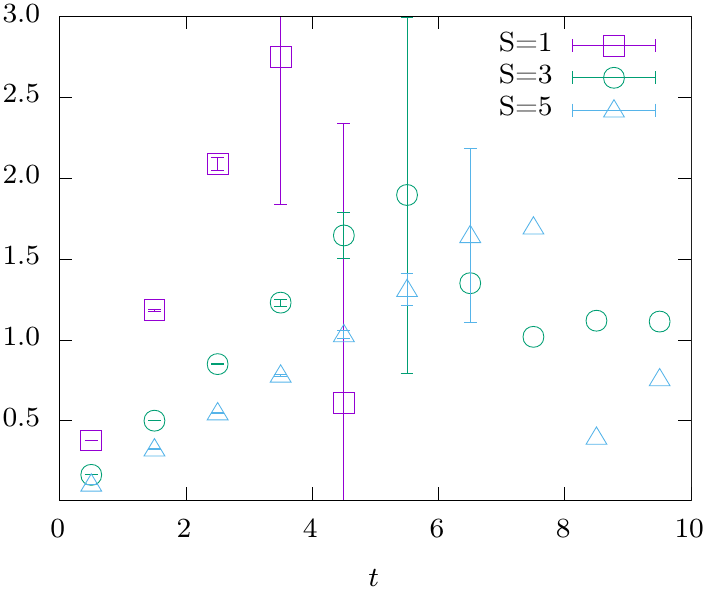}	
  \inputFig{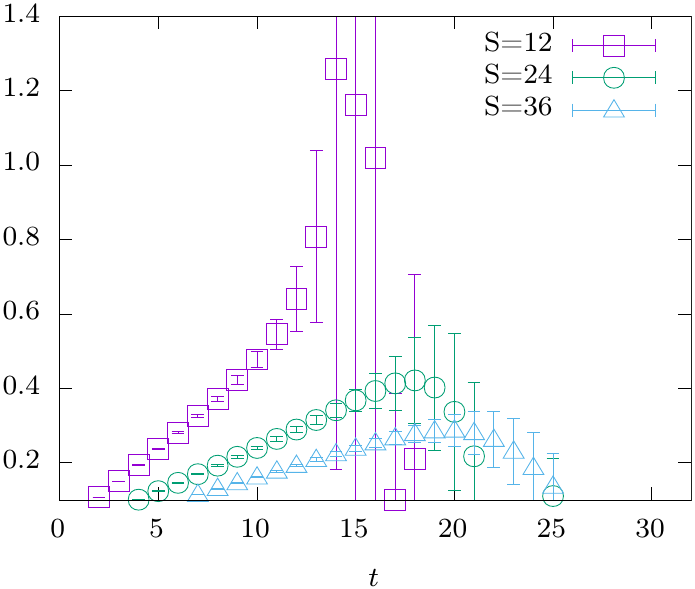}
\caption{Comparison between the glueball correlation function $C(t)$ as function of the temporal extend $t$ for the two-dimensional Yang-Mills theory (left) and the two-dimensional Super Yang-Mills theory (right) for different smearing levels $S$. In the bottom row we plot the effective mass.}
\label{fig::YMvsSYM}
\end{figure}
The only nonzero values of the correlation function are at distances $t=0,1,63$ and $64$. A similar behavior is seen in pure Yang-Mills theory on a 
two-dimensional lattice. Indeed, with \mbox{Migdals prescription \cite{Migdal:1975zg}} one
obtains for the correlation function of the glueball operator $G(x)$ in this theory
\begin{equation}
\langle G(x)G^\dagger(y)\rangle=C_\text{G}=\text{const.}
\end{equation}
This holds true in case the supports of the interpolating operators
are disjunct.
Hence the correlation function of glueballs will show only a correlation 
between time slices with distance less than the diameter $A_G$ of
the support of $G(x)$. We observe the very same behaviour in the supersymmetric theory 
in Figure~\ref{fig::glueballcorr}, where the diameter is two. 
In the continuum limit, the physical diameter shrinks to zero and the expectation value is constant in the whole spacetime volume. Furthermore one can show, that this value goes to zero and the glueball decouples completely from the theory. 
This lattice result is in agreement with the analytical result presented in \cite{Bralic:1980ra}. 

Since we use smearing of sources and sinks in our analysis, it maybe
instructive to study the effect of smearing on the correlation 
function of glueballs. Every smearing step increases the diameter $A_\text{G}$, 
and thus induces more artificial correlations between the lattice points, 
which are uncorrelated without smearing.
The results can be seen in Figure~\ref{fig::YMvsSYM}, where we compare pure 
Yang-Mills theory (left) to susy Yang-Mills theory (right).
In both cases we observe more nonzero values in 
the correlation functions for higher smearing levels, as expected. 
Smearing effects can also be seen in the effective mass:
in both theories it is an ever increasing function 
of the distance for all values of the smearing level. We conclude that,
similarly as in pure YM-theory in two dimensions, there is no correlation for glueballs. In other words, the glueball completely decouples 
from the $\mathcal{N}=(2,2)$ SYM theory in two dimensions. 
Similarly we could not detect any correlations in the scalarball 
and glue-scalarball correlator functions.
Since they should form a super-multiplet with the glueball,
they will decouple from the theory as well. The additional gluino-glueball state in the super-multiplet will also show no correlations, 
and thus is not seen in our simulations.


\section{Conclusions}
\label{sec::conclusions}

In our work, we simulated the two-dimensional $\sN=(2,2)$ SYM lattice-theory
in a conventional approach without twisting. The simulation could
be afflicted with two potentially serious problems common in gauge theories
with extended supersymmetry: flat directions and a sign problem. 
In the present work we demonstrate that these problems do not arise 
for all parameters which are relevant to approach the supersymmetric continuum limit. 
As concerning the sign problem, this is related to the absence of the sign problem 
in the $\mathcal{Q}$-exact formulation of the continuum theory 
\cite{Catterall:2011aa}.

When studying various Ward identities, we did observe that they are rather
insensitive to the bare mass of the scalars $m_\text{s}$, as long as the latter
is in the vicinity of the (all-loop) perturbative value in the
supersymmetric continuum model, which is given by \mbox{$m_\text{s}^2=0.659\,482\,55(8)$}. Away from the continuum
limit this may not be the optimal choice. Spotting an observable, which 
allows for further fine-tuning of the scalar mass on the lattice
could perhaps improve the results and would allow for more accurate 
predictions. But such an improvement is probably not easy to achieve since our 
results are stable and reliable. They do not depend 
on the scalar mass in the vicinity of the above value and thus
a further fine-tuning of $m_\text{s}$ does not help much.

The restoration of supersymmetry is observed in the chiral limit.
Since the fermion mass is not a relevant coupling
(contrary to the situation in four
dimensions) this may come as a surprise.
But generally speaking fine-tuning of an irrelevant coupling may be helpful away 
from criticality. In any case, the result confirms the assumption, 
that supersymmetry is recovered in the chiral limit, similarly
as in the four-dimensional mother-theory.
But the spectrum of bound states looks different than in the four-dimensional 
$\mathcal{N}=1$ theory. We found a massless multiplet -- the dimensionally reduced Veneziano-Yankielowicz multiplet -- which contains the mesons, while 
the Farrar-Gabadadze-Schwetz multiplet decouples from the theory (see Table~\ref{tab::finalMass}). 
\begin{table}[ht]
\begin{center}
\begin{tabular}{c|c|c}
\toprule[1pt]
particle & $m$ & $m^*$  \\ \midrule[0.5pt]
a-$\eta$ & $0.034(2) \to 0$ & $0.278(25)$ \\
a-f & $0.034(1) \to 0$ & $0.433(4)$ \\
gluino-glueball & -- & $0.243(3)$/$0.287(10)$ \\
\bottomrule[1pt]
\end{tabular}
\end{center}
\caption{We observe the formation of a massive VY-multiplet while the ground states are massless. The FGS-multiplet decouples from the theory.\label{tab::finalMass}}
\end{table}
The mass of the lightest gluino-glueball seen in the simulations
is still a bit ambiguous. Within errors its mass is equal to that
of the excited mesons. 
We believe we could not follow the corresponding correlation function
for large enough $t$-values, in order to disentangle the signals 
from the ground state and excited state. Probably we did only
see the excited gluino-glueball which forms a multiplet with the 
excited meson states. If this is true, then finding the missing
ground state of the gluino-glueball may be as difficult as finding
a needle in a haystack.

In this work we could not see any screening of static
charges in the fundamental representation, although
the dynamical fermions are in the adjoint. Instead our 
accurate simulations indicate that $\mathcal{N}=(2,2)$ and $\mathcal{N}=(1,1)$ SYM theory in two
dimensions both confine static charges in
the fundamental representation.
At least the result for the $\mathcal{N}=(1,1)$ theory with
Majorana fermions seems to be in 
conflict with analytic results in \cite{Gross:1995bp}.
Clearly, this clash of numerical simulations with analytical
results should be resolved in future works.

In future studies we intend to study the phase structure of the
$\mathcal{N}=(2,2)$ SYM theory as well as related systems with more
supersymmetries. It would be interesting to measure
the two independent holonomies (Wilson loops with windings) 
on the two-torus and their dependence on the geometry of the torus.
This way one could first compare with results obtained with $\mathcal{Q}$-exact
formulations for $\mathcal{N}=(8,8)$ SYM theory \cite{Catterall:2017lub}
and furthermore extend 
to systems with less supersymmetry where no $\mathcal{Q}$-exact formulation exists.
Since we did not encounter any sign problems for $\kappa<\kappa_c$ and since
the flat directions are stabilized, we should be able to accurately
localize the expected phases and phase-transition lines in two-dimensional SYM with extended supersymmetry.

\acknowledgments
We thank Martin Ammon, Georg Bergner and Masanori Hanada
for fruitful discussions and comments. This 
work was supported by the DFG Research Training Group 1523 “Quantum and Gravitational Fields” and in part by the DFG-Grant Wi777/11. The simulations were performed at the HPC-Clusters OMEGA and ARA of the University Jena.
\appendix

\section{Exact lattice Ward identities}\label{Appendix_A}
In the main body of the text we studied the violation of
several Ward identities due to lattice artifacts. Thereby we neglected
contributions stemming from $m_\text{f}$ and $m_\text{s}$
deviating from their critical values. Here we derive
lattice Ward identities without any approximation.
The application of the lattice supersymmetry 
transformations \eqref{eq:latticeSusyTrafo} to the lattice Lagrangian results in
\begin{equation}
\begin{aligned}
	\bar{Q}^\alpha \mathcal{L}_\text{lat}
 =&\frac{\beta}{2}\left\lbrace{\partial_\mu s_\mu}^\alpha -2 m_\text{f}\,{(\Gamma_{MN})^\alpha}^\beta F^{MN}\lambda_\beta\right\rbrace+2 m_\text{s}^2 \,{(\Gamma_{m+1})^\alpha}_\beta\lambda^\beta \phi^m+X_\text{S}\\
=&\frac{\beta}{2}\left\{\partial_\mu s_\mu^\alpha 
- m_\text{f}\,\chi_\text{f}^\alpha \right\}
+ m^2_\text{s}\,\chi_\text{s}^{\alpha}+X_\text{S},
\end{aligned}
\end{equation}
with
\begin{equation}
 \chi_\text{f}^\alpha=2 \tr\big(\Gamma_{MN}^{\alpha\beta} F^{MN}\lambda_\beta\big)\quad \text{and} \quad
 \chi_\text{s}^\alpha=2 \tr\big(\Gamma_{m+1}^{\alpha\beta}\lambda_\beta \phi^m\big)\,.
\end{equation}
The contributions $\chi^\alpha$ originate from the fermion and scalar mass 
terms introduced in the lattice Lagrangian. 
As pointed out previously the supercurrent $s_\mu^\alpha$ vanishes after 
summation over the lattice sites. The term $X_\text{S}$ originates 
from the lattice regularisation and is of order $\mathcal{O}(a)$. 
Clearly, at tree-level supersymmetry is restored in the continuum limit for
the critical values $m_\text{f}^\text{c}=m_\text{s}^\text{c}=0$.
At one-loop a finite scalar mass is generated due to different 
lattice momenta of bosons and fermions. 
Furthermore, the Wilson term in the fermion operator gives
rise to a nonzero critical fermion mass.
In the continuum limit, no further corrections are generated at higher 
loop order such that $m_\text{f}^\text{c}\to 0$.
In order to compensate for the shifts at finite lattice spacing one adds
counter-terms to the tree-level lattice action and ends up with the full 
quantum lattice Ward identity \eqref{lat_wid2}. The scalar mass counter-term 
must also be included in the Ward identity $W_3$ and the bosonic Ward 
identity because they contain the kinetic term for the scalar fields. 
Thus, the set of lattice Ward identities read
\begin{equation}
 \begin{aligned}
	W_\text{B}=&\beta V^{-1}\langle S_\text{B}\rangle
	+ m_\text{s}^2 \langle\tr\phi^2\rangle 
	+\beta\langle \tr \bar{\lambda}\,\Gamma^{MN}F_{MN}\,\Theta\rangle
	\to\frac{9}{2}\,, \\
	W_3=&\frac{\beta}{2} \langle \tr D_\mu\phi^a D^\mu\phi_a\rangle
	+  m_\text{s}^2 \langle\tr\phi^2\rangle 
	+2\beta\,\big\langle \tr\bar{\lambda}\,\Gamma^{\mu m}D_\mu \phi_m\,\Theta\big\rangle	
	\to 3\,, \\
	W_2=&\frac{\beta}{4}\langle \tr F_{\mu\nu}F^{\mu\nu}\rangle
	+\beta\langle\tr\bar{\lambda}\Upsilon\big\rangle 
	+\beta\langle\tr\bar{\lambda}\,\Gamma^{\mu\nu}F_{\mu\nu}\Theta \rangle
	\to \frac{3}{2}\,, \\
	W_1=&\frac{\beta}{2}\langle\tr\left[\phi_1,\phi_2\right]^2\rangle
	-\beta\langle\tr\bar{\lambda}\,\Upsilon\rangle 
	+\beta\langle\tr\bar{\lambda}\,\Gamma^{mn}\left[\phi_m,\phi_n\right]\Theta\rangle
	\to 0\,,\label{app::fullward}
 \end{aligned}
\end{equation}
where we used the abbreviations 
\begin{equation}
\Theta=\left(m^2_\text{s}-\left(m^\text{c}_\text{s}\right)^2\right)\chi_\text{s}-\left(m_\text{f}-m^\text{c}_\text{f}\right)\chi_\text{f},\quad
\Upsilon=\frac{\ii}{8}\big(
\Gamma_2\left[\phi_1,\lambda\right]+\Gamma_3\left[\phi_2,\lambda\right]\big)\,.
\end{equation}
Near the supersymmetric continuum limit, lattice artifacts should be 
sufficiently suppressed such that the breaking of Ward identities originate
from the missing fine-tuning of
$m_\text{f}^c$ and $m^\text{c}_\text{s}$. Since we anyway use
the $\pi$-mass to fine-tune $m_\text{f}^c$ we will focus 
on the fine-tuning of $m^\text{c}_\text{s}$ in what follows.
We will show this fine-tuning approach for the Ward-identity $W_2$. 
The results for the other identities are very similar.

First we introduce $W_2^\text{b}$ and the correction terms $\mathcal{C}_\text{s}$ and $\mathcal{C}_\text{f}$ 
\begin{equation}
	W_2^\text{b}=\beta\big\langle \frac{1}{4}\tr F_{\mu\nu}F^{\mu\nu}+\tr\bar{\lambda}\Upsilon\big\rangle,
	\quad 
	\mathcal{C}_\text{s}=\langle\tr\bar{\lambda}\,\Gamma^{\mu\nu}F_{\mu\nu}\chi_\text{s}\rangle,
	\quad 
	\mathcal{C}_\text{f}=\beta\langle\tr\bar{\lambda}\,\Gamma^{\mu\nu}F_{\mu\nu}\chi_\text{f}\rangle \,,
\end{equation}
which enter the Ward identity $W_2$ of interest,
\begin{equation}
W_2=W_2^\text{b}+\left(m^2_\text{s}-\left(m^\text{c}_\text{s}\right)^2\right)\mathcal{C}_\text{s}+\left(m_\text{f}-m^\text{c}_\text{f}\right)\mathcal{C}_\text{f} \,.
\label{appw1}
\end{equation}
Now we simulate the gauge theory for a set of values
$m^2_\text{s}$ near the
one-loop value $0.6594826$ and measure
the expectation values $W_2^\text{b}$, $\mathcal{C}_\text{s}$ 
and $\mathcal{C}_\text{f}$.
Note that $m_\text{s}$ and $m_\text{f}$ are the masses used to
generate the ensemble, whereas the trial mass
$m^{\text{c}}_\text{s}$ only enters via the operators defining 
the Ward identities.
Next we should extract a trial mass
for which $W_2\approx \frac{3}{2}$ for all $m_\text{s}$ near the 
critical value. Note that the extracted $m^\text{c}_\text{s}$ could 
deviate from the one-loop results due to lattice artifacts.

\begin{figure}[htb]
\inputFig{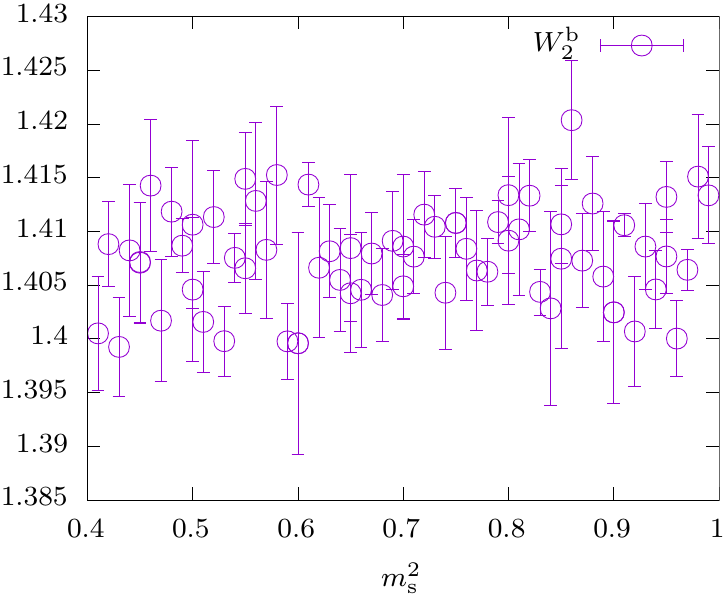}\hskip5mm
\inputFig{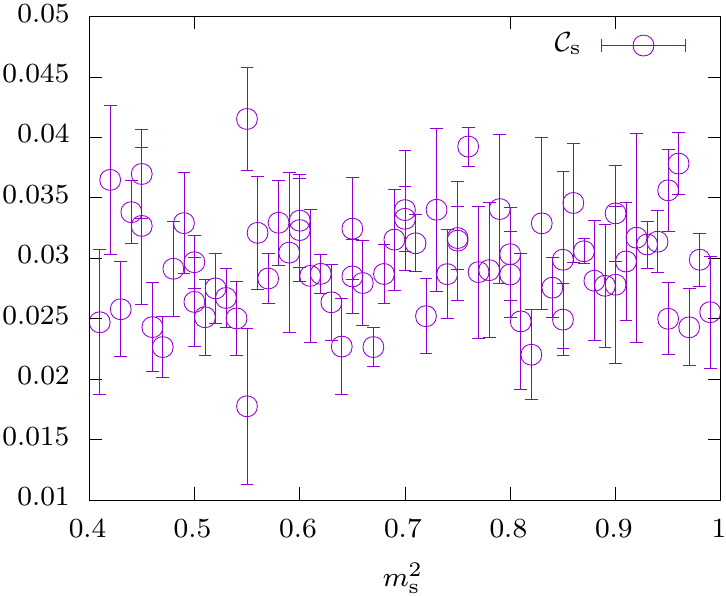}
\caption{On the left we see the term $W_2^\text{b}$ and on the 
right the term $\mathcal{C}_\text{s}$. \label{fig::appendix}}
\end{figure}

Figure~\ref{fig::appendix} clearly shows that $W_2^\text{b}$
and $\mathcal{C}_\text{s}$ do not depend sensitively on $m_\text{s}$
near the critical one-loop value. The same holds true 
for $\mathcal{C}_\text{f}$, which is not shown in the figure. 
This means that it is difficult to find any deviations  
of $m^\text{c}_\text{s}$  from its known continuum one-loop value.
But since the correction terms $\mathcal{C}_\text{s}$ and 
$\mathcal{C}_\text{f}$ in (\ref{appw1}) are 
two orders of magnitude smaller than $W_2^\text{b}$ we may safely 
neglect the  lattice correction $\Theta$ if we are close to the 
critical masses, which we ensure by extrapolating to the chiral limit and 
using $0.6594826$. 
This leads to the final  set of approximate Ward identities (\ref{ward_dec}) which
are measured in our simulations.

\section{Meson correlation functions}\label{Appendix_B}
In order to extract meson masses, we measure the connected two-point functions of the operators $\bar{\lambda}_m \Gamma \lambda_n$,
\begin{equation}
 C_{\Gamma,m,n}(x,y)=\erw{\left(\bar{\lambda}_m \Gamma \lambda_n\right)_{x}\left(\bar{\lambda}_n \Gamma \lambda_m\right)_{y}}-\erw{\left(\bar{\lambda}_m \Gamma \lambda_n\right)_{x\vphantom{y}}}\erw{\left(\bar{\lambda}_n \Gamma \lambda_m\right)_{y}}
\end{equation}
with $\Gamma=\id$ for the scalar mesons and $\Gamma=\Gamma_5$ for the axial mesons. The indices $m,n$ are flavour indices. In a two-flavour setup, the f-meson mass is extracted from the decay of
$C_\text{f}=C_{\id,1,1}$, the $\eta$-meson mass from $C_\eta=C_{\Gamma_5,1,1}$ and the pion mass from $C_\pi=C_{\Gamma_5,1,2}$. After integration over the fermions, we obtain
\begin{equation}
\begin{aligned}
 C_{\Gamma,m,n}(x,y)=&\erw{\tr\left(\Delta_{x,x}^{mn}\Gamma\right)\tr\left(\Delta_{y,y}^{nm}\Gamma\right)-\tr\left(\Delta_{x,y}^{mm}
 \Gamma\Delta_{y,x}^{nn}\Gamma\right)}\\
 &-\erw{\tr\left(\Delta^{mn}_{x,x}\Gamma\right)}\erw{\tr\left(\Delta^{nm}_{y,y}\Gamma\right)}
\end{aligned}
\end{equation}
with the fermion propagator $\Delta$. In our simulations, only one fermion flavour is dynamic. The pion correlation function is therefore defined in a \emph{partially quenched} setup which implies $\Delta^{11}=\Delta^{22}=\Delta$ and $\Delta^{1,2}=\Delta^{2,1}=0$.
We get for the different correlation functions
\begin{equation}
\begin{aligned}
 C_\text{f}(x,y)=&\erw{\tr\left(\Delta_{x,x}\right)\tr\left(\Delta_{y,y}\right)-\tr\left(\Delta_{x,y}\Delta_{y,x}\right)}-\erw{\tr\left(\Delta_{x,x}\right)}\erw{\tr\left(\Delta_{y,y}\right)}\,,\\
 C_\eta(x,y)=&\erw{\tr\left(\Delta_{x,x}\Gamma_5\right)\tr\left(\Delta_{y,y}\Gamma_5\right)-\tr\left(\Delta_{x,y}\Gamma_5\Delta_{y,x}\Gamma_5\right)}-\erw{\tr\left(\Delta_{x,x}\Gamma_5\right)}\erw{\tr\left(\Delta_{y,y}\Gamma_5\right)}\,,\\
 C_\pi(x,y)=&\erw{-\tr\left(\Delta_{x,y}\Gamma_5\Delta_{y,x}\Gamma_5\right)}.
\end{aligned}
\end{equation}
For a single flavour, the pion correlation function is therefore defined as the \emph{connected} part of the $\eta$-meson correlation function, where connected refers to a diagramatical interpretation of traces over the fermion propagator.

\bibliographystyle{JHEP}
\bibliography{paper.bib}

\end{document}